\documentclass[sigconf]{acmart}

\usepackage{lipsum}								

\usepackage[utf8]{inputenc}                     
\usepackage[TS1,T1]{fontenc}			        

\usepackage{xpatch}								

\usepackage[babel]{csquotes}					

\usepackage{booktabs}							

\usepackage{amsmath}							
\usepackage{amsfonts}							

\usepackage[disable]{todonotes}		            

\usepackage{listings, lstautogobble}			

\usepackage[mathscr]{eucal}						

\usepackage{colortbl}							

\usepackage{amsthm}								

\usepackage[ruled]{algorithm}					
\usepackage{algorithmic}						

\usepackage{bookmark,hyperref}					

\AtBeginDocument{%
	}
		
\newcommand\numberstyle[1]{%
	\footnotesize
	\color{SQLcodegray}%
	\ttfamily
	\ifnum#1<10 0\fi#1 |%
}

\makeatletter
\xpatchcmd\algorithmic
{\ALC@linenosize \arabic{ALC@line}\ALC@linenodelimiter}
{\ALC@linenosize \theALC@line\ALC@linenodelimiter}
{\ALC@linenosize \tiny}
{}{}

\newcommand\setAlgoLinenoFormat{\renewcommand*{\theALC@line}{\ifnum\value{ALC@line}<10 0\fi\arabic{ALC@line}}}
\makeatother

\setAlgoLinenoFormat

\let\oldalgorithmic\algorithmic
\renewcommand\algorithmic{\ttfamily\scriptsize\fontseries{l}\selectfont\oldalgorithmic}

\MakeAutoQuote{«}{»}

\definecolor{aliceblue}{rgb}{0.94, 0.97, 1.0}
\definecolor{skyblue}{rgb}{0.53, 0.81, 0.92}
\definecolor{brightmaroon}{rgb}{0.95, 0.82, 0.85}

\hypersetup{%
	breaklinks=true,			
	colorlinks=true,			
	urlcolor=blue,				
	citecolor=gray,				
	linkcolor=darkgray,			
	anchorcolor=blue,			
	bookmarksopen=false,		
	linktocpage=true,			
	bookmarksnumbered,			
	pdfstartview={FitH},		
	pdftitle={Cubes de données hiérarchiques},
	pdfauthor={Mickaël Martin Nevot, Sébastien Nedjar, Lotfi Lakhal},
	pdfsubject={Article de recherche},
	pdfkeywords = {ROLAP cubing, Data warehouse, Datacube, Big data, Business intelligence, Hierarchical cube, Hierarchical dimensions},
}

\definecolor{SQLCodeGreen}{rgb}{0,0.6,0}
\definecolor{SQLcodegray}{rgb}{0.5,0.5,0.5}
\definecolor{SQLCodePurple}{HTML}{C42043}
\definecolor{SQLBackgroundcolor}{HTML}{F2F2F2}
\definecolor{SQLBookColor}{cmyk}{0,0,0,0.90}  
\color{SQLBookColor}

\lstdefinestyle{SQLStyle} {
	basicstyle=\tiny
	backgroundcolor=\color{SQLBackgroundcolor},
	commentstyle=\color{SQLCodeGreen},
	keywordstyle=\color{SQLCodePurple},
	numberstyle=\numberstyle,
	stringstyle=\color{SQLCodePurple},
	basicstyle=\footnotesize\ttfamily,
	breakatwhitespace=false,
	breaklines=true,
	captionpos=b,
	keepspaces=true,
	numbers=left,
	numbersep=10pt,
	showspaces=false,
	showstringspaces=false,
	showtabs=false,
	autogobble=true,
	literate = 	{é}{{\'e}}{1}%
				{è}{{\`e}}{1}%
				{à}{{\`a}}{1}%
				{â}{{\^a}}{1}
				{ç}{{\c{c}}}{1}%
				{œ}{{\oe}}{1}%
				{ù}{{\`u}}{1}%
				{É}{{\'E}}{1}%
				{È}{{\`E}}{1}%
				{À}{{\`A}}{1}%
				{Ç}{{\c{C}}}{1}%
				{Œ}{{\OE}}{1}%
				{Ê}{{\^E}}{1}%
				{ê}{{\^e}}{1}%
				{î}{{\^i}}{1}%
				{ï}{{\"i}}{1}
				{ô}{{\^o}}{1}%
				{û}{{\^u}}{1}%
}

\newtheoremstyle{theoremStyle}
{9pt}{9pt}				
{}						
{}						
{\bfseries}{ -}			
{ }						
{}						

\theoremstyle{theoremStyle}

\pgfdeclarelayer{background}
\pgfdeclarelayer{foreground}
\pgfsetlayers{background, main, foreground}

\usetikzlibrary{patterns}
\usetikzlibrary{automata}
\usetikzlibrary{angles}
\usetikzlibrary{quotes}

\floatname{algorithm}{Algorithme}

\renewcommand{\algorithmicif}{\textbf{si}}
\renewcommand{\algorithmicthen}{\textbf{alors}}
\renewcommand{\algorithmicelse}{\textbf{sinon}}

\renewcommand{\algorithmicfor}{\textbf{pour}}
\renewcommand{\algorithmicforall}{\textbf{pour tout}}
\renewcommand{\algorithmicdo}{\textbf{faire}}

\renewcommand{\algorithmicloop}{\textbf{boucle}}

\newcommand{\LET}{\STATE \algorithmiclet\ }
\newcommand{\INPUT}{\item[\algorithmicinput]}
\newcommand{\OUTPUT}{\item[\algorithmicoutput]}
\newcommand{\algorithmiclet}{\textbf{soit}}
\newcommand{\algorithmicinput}{\textbf{Entrée \string:}}
\newcommand{\algorithmicoutput}{\textbf{Sortie \string:}}
\newcommand{\algorithmicexit}{\textbf{exit}}

\makeatletter
\newcommand{\FORALE}[2][default]{\ALC@it\algorithmicforall\textbf{e}\ #2\ \algorithmicdo\ALC@com{#1}\begin{ALC@for}}
\newcommand{\EXITFOR}{\ALC@it\algorithmicexit\ \algorithmicfor}
\newcommand{\EXITLOOP}{\ALC@it\algorithmicexit\ \algorithmicloop}
\newcommand{\FORALLOL}[2]{\ALC@it\algorithmicforall\ #1\ \algorithmicdo\ #2}
\newcommand{\IFTHENOL}[2]{\ALC@it\algorithmicif\ #1\ \algorithmicthen\ #2 \\}
\newcommand{\IFTHENELSEOL}[3]{\ALC@it\algorithmicif\ #1\ \algorithmicthen\ #2\ \algorithmicelse\ #3}
\newcommand{\THENOLO}[1]{\ALC@it\hspace*{2,15cm}\algorithmicelse\ #1\\}
\newcommand{\THENOLA}[1]{\ALC@it\hspace*{2,65cm}\algorithmicelse\ #1\\}
\newcommand{\THENOLB}[1]{\ALC@it\hspace*{1,85cm}\algorithmicelse\ #1\\}
\newcommand{\THENOLC}[1]{\ALC@it\hspace*{1,95cm}\algorithmicelse\ #1\\}
\makeatother

\ifdefined\proposition\else
    \newtheorem{proposition}{Proposition}[section]

    \ifdefined\lemma\else
        \newtheorem{lemma}[proposition]{Lemme}
    \fi

    \ifdefined\corollaire\else
        \newtheorem{corollaire}[proposition]{Corollaire}
    \fi

\fi

\ifdefined\lemma\else
    \newtheorem{lemma}{Lemme}
\fi

\ifdefined\corollaire\else
    \newtheorem{corollaire}{Corollaire}
\fi

\ifdefined\theorem\else
    \newtheorem{theorem}{Théorème}
\fi

\ifdefined\definition\else
    \newtheorem{definition}{Definition}[section]
\fi

\ifdefined\example\else
    \newtheorem{example}{Exemple}[section]
\fi

\ifdefined\property\else
    \newtheorem{property}{Property}[section]
\fi

\ifdefined\remarques\else
    \newtheorem{remarques}{remarques}
\fi

\setcopyright{none}
\copyrightyear{2024}
\acmConference[40ème conférence sur la gestion des données (BDA)]{BDA}{October 21-24, 2024}{Orléans, France}

\acmISBN{978-1-4503-XXXX-X/18/06}

\begin{document}
	
	\ifdefined\listoftodos
	\listoftodos
	\fi
	
	
	\title{Classement d'objets Skylines dans les bases de données}

\author{Mickaël Martin Nevot}
\email{mickael.martin-nevot@lis-lab.fr}
\orcid{0009-0004-7893-3449}
\affiliation{%
	\institution{Aix-Marseille Université LIS CRNS UMR 7020}
	\city{Marseille}
	\country{France}
}

\author{Lotfi Lakhal}
\email{lotfi.lakhal@lis-lab.fr}
\affiliation{%
	\institution{Aix-Marseille Université LIS CRNS UMR 7020}
	\city{Marseille}
	\country{France}
}





\renewcommand{\shortauthors}{Mickaël Martin Nevot et Lotfi Lakhal}

\begin{abstract}
	L'analyse multicritère dans les bases de données a été activement étudiée, en particulier avec l'utilisation de l'opérateur Skyline. Pourtant, peu d'approches proposent un classement pertinent des points Pareto-optimal, ou Skyline, permettant d'ordonner les résultats à forte cardinalité. Nous proposons d'améliorer la méthode dp-idp, inspiré de tf-idf, une approche récente attribuant un score à chaque point du Skyline, en introduisant le concept de hiérarchie de dominance. Comme dp-idp ne garantit pas un classement distinctif, nous introduisons la méthode CoSky, de type TOPSIS, issue à la fois de la recherche d'information et de l'analyse multicritère. CoSky, intégrable directement dans un SGBD, effectue une pondération automatique d'attributs normalisés grâce à l'indice de Gini, suivi d'un calcul de score avec le cosinus de Salton par rapport à un point idéal déterminé. En couplant le principe de Skyline multiniveaux à CoSky, nous introduisons l'algorithme DeepSky. La mise en œuvre des méthodes dp-idp et CoSky sont évaluées expérimentalement.
\end{abstract}

%




\maketitle
	
	
	\section{Introduction}

L'opérateur Skyline (\cite{borzsonyiSkylineOperator2001}), précédemment ensemble de Pareto et vecteurs maximaux (\cite{bentleyAverageNumberMaxima1978}), est capital dans l'analyse multicritère et a été largement étudié. Ses principales problématiques sont une forte cardinalité et une faible corrélation dans un jeu de données. Dans ces cas-là, il est souvent difficile d'extraire d'information significative car plus les points d'un Skyline sont nombreux plus ils peuvent avoir des intérêts proches et de faibles différences significatives. Un classement efficace permet d'y pallier, mais peu de telles approches ont été proposées bien qu'elles permettent de faciliter et prioriser la prise de décision, réduire la complexité de l'analyse, simplifier l'évaluation de compromis et proposer des solutions en fonction de préférences spécifiques ou d'objectifs contextuels.

L'approche dp-idp est certainement une des approches les plus récentes. Elle utilise la dominance de Pareto afin de déterminer un score pour chaque point d'un Skyline. dp-idp reprend l'idée du schéma de pondération tf-idf utilisé en recherche d'information \footnote{La recherche d'information (RI, ou IR pour \emph{information retrieval}), englobe la recherche et la récupération de données ou d'informations pertinentes à partir de données non structurées, telles que des textes, des images, des vidéos ou des sons.}.

Dans ce papier, nous proposons tout d'abord d'améliorer la méthode dp-idp en utilisant le concept de hiérarchie de dominance afin de perfectionner le calcul des scores des points d'un Skyline, puis nous définissons la méthode CoSky afin de classer efficacement les points d'un Skyline sans privilégier de dominance, avant de présenter l'algorithme DeepSky, un algorithme Skyline multiniveaux utilisant la méthode CoSky pour classer les top-$k$ points de Skyline. Enfin, la mise en œuvre des méthodes dp-idp et CoSky sont présentées avec des évaluations expérimentales.

Ce papier est une révision approfondie du travail initial (\cite{alouaouiCoSkyPracticalMethod2019}). Nos principaux nouveaux apports portent sur la spécification et l'implémentation des méthodes proposées~: dp-idp avec hiérarchie de dominance et CoSky, leurs évaluations expérimentales confrontées à celle de l'algorithme de référence SkyIR-UBS et leurs commentaires, l'unification des préférences Skyline ainsi que la clarification, la précision et l'amélioration de la méthode CoSky.

\section{Cas d'utilisation}

Nous considérons ici un exemple appliqué au jeu vidéo Pokémon Showdown!\footnote{Pokémon Showdown! est un jeu vidéo sur navigateur Web et PC \emph{open source}. C'est un simulateur de combat de Pokémon (Pokémon ne prend pas de marque du pluriel, possiblement car il s'agit d'un terme issu du japonais) populaire (des millions d'utilisateurs mensuels, avec jusqu'à plus de 20000 simultanément) permettant de jouer à des combats de Pokémon en ligne animés.}, et sa relation exemple \texttt{Pokémon} (cf. tableau~\ref{tab:relation_exemple}).

\begin{table}[htb]
	\caption{La relation \texttt{Pokémon}}\label{tab:relation_exemple}
	\centering
	\begin{minipage}{\linewidth}
		\resizebox{1\textwidth}{!}{
			\begin{tabular}{c|cc|ccc}
				\toprule
				\texttt{RowId} & \texttt{Joueur}\footnote{Avec les numéros officiels des Pokémon~: n°065~: Alakazam, n°080~: Flagadoss, n°103~: Noadkoko, n°113~: Leveinard, n°121~: Staross, n°128~: Tauros, n°143~: Ronflex.} & \texttt{Adversaire}\footnote{\emph{idem supra}.} & \texttt{Rareté}\footnote{Soit $p$ le pourcentage d'obtention d'une séquence de Pokémon (qui est la multiplication du pourcentage d'obtention de chaque Pokémon de la séquence), le score de \texttt{Rareté} $r$ est calculé, sur une échelle allant de 0 à 10, de la manière suivante~: $\text{si } p = 1 \text{ alors } r = 0 \text{, sinon } r = \lfloor max(\frac{(p - 1) \times 10 - (100 - e \times 0.9)}{e}, 0) \rfloor + 1 $.} & \texttt{Durée}\footnote{En nombre de tours de combat au total.} & \texttt{Victoire}\footnote{En pourcentage.} \\
				\midrule
				$1$ & $121, 113, 103$ & $121, 113, 121$ & $5$ & $20$ & $70$ \\
				$2$ & $065, 103, 065$ & $065, 143, 065$ & $4$ & $60$ & $50$ \\
				$3$ & $121, 113, 121$ & $065, 103, 065$ & $5$ & $30$ & $60$ \\
				$4$ & $121, 113, 080$ & $065, 143, 065$ & $1$ & $80$ & $60$ \\
				$5$ & $121, 113, 128$ & $121, 113, 121$ & $5$ & $90$ & $40$ \\
				$6$ & $065, 113, 143$ & $065, 113, 143$ & $9$ & $30$ & $50$ \\
				$7$ & $065, 143, 065$ & $121, 113, 143$ & $7$ & $80$ & $60$ \\
				$8$ & $065, 113, 143$ & $065, 103, 065$ & $9$ & $90$ & $30$ \\
				\bottomrule
			\end{tabular}
		}
	\end{minipage}
\end{table}

\begin{figure}[htb]
	\begin{minipage}{\linewidth}
		\centering
		\includegraphics[width=\linewidth]{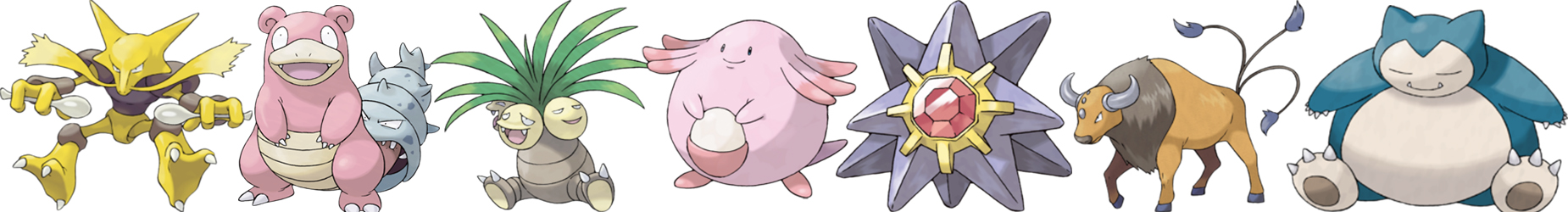}
		\caption[Alakazam, Flagadoss, Noadkoko, Leveinard, Staross, Tauros, Ronflex]{Les Pokémon du cas d'utilisation\footnote{De gauche à droite~: Alakazam (n°065), Flagadoss (n°080), Noadkoko (n°103), Leveinard (n°113), Staross (n°121), Tauros (n°128), Ronflex (n°143)~; illustrations de Pokémon Versions Rouge Feu et Vert Feuille sur Poképédia.}}\label{fig:alakazam_flagadoss_noadkoko_leveinard_staross_tauros-ronflex}
	\end{minipage}
\end{figure}

\section{Définitions préliminaires}

\subsection{Préférence et dominance}\index{Préférence Skyline}\index{Dominance Skyline}

Soit $r$ une relation avec des attributs $A_1, \dotsc,  A_m$. Une préférence Skyline sur $A_j$ est une expression de l'une des deux formes suivantes~:  $Pref(A_i) = \texttt{MIN}$, ou $Pref (A_i) = \texttt{MAX}$. Une préférence Skyline décrit donc les situations préférables. Soit $t$ et $t'$ deux tuples de $r$. Nous considérons que $t$ domine $t'$ (noté $t \prec_d t'$) si et seulement si $t[A_1] \le t'[A_1], \dotsc, t[A_m] \le t'[A_m]$ et $\exists j \in [1..m] : t[A_j] < t'[A_j]$ avec~:
\begin{equation}
	(\preceq_d, \prec_d) =
	\left\lbrace
	\begin{array}{l}
		(\le, <) \equiv Pref(A_j) = \texttt{MIN} \\ 
		(\ge, >) \equiv Pref(A_j) = \texttt{MAX}
	\end{array}
	\right.
\end{equation}

\begin{example}
	La relation \texttt{Pokémon} illustrée par le tableau~\ref{tab:relation_exemple} est typique pour l'utilisation de calcul de Skyline. L'attribut \texttt{Joueur} est la séquence de Pokémon jouée par le joueur, et \texttt{Adversaire}, celle de l'adversaire. Les critères déterminant le «~meilleur ordre d'apparition de Pokémon dans un combat~» sont la \texttt{Rareté} des Pokémon de la séquence jouée par le joueur, la \texttt{Durée} (en nombre de tours) du combat et le taux de \texttt{Victoire} de la séquence de Pokémon du joueur par rapport à celle de l'adversaire. \texttt{Rareté} et \texttt{Durée} sont des critères à minimiser, alors que \texttt{Victoire} est à maximiser (les préférences Skyline considérées sont donc mixtes). Pour l'exemple, nous nous limitons à trois Pokémon par séquence, qu'elle soit du joueur ou de l'adversaire, et afin de simplifier leurs usages dans leurs représentations nous les indiquons par des listes de numéros de Pokémon, et leur affectons une lettre. Ainsi, la liste $121, 113, 128 (A)$ correspond à la séquence des trois Pokémon Staross, Leveinard et Tauros. Pour la lisibilité, nous arrondissons à un multiple de cinq \texttt{Durée} et \texttt{Victoire}.
\end{example}

\subsection{Opérateur Skyline}\index{Opérateur Skyline}

Une syntaxe SQL de l'opérateur Skyline a été proposée pour exprimer des requêtes basées sur des préférences (\cite{borzsonyiSkylineOperator2001}).

Dans l'exemple, la requête SQL avec opérateur Skyline est~:

\lstset{style=SQLStyle}
\begin{lstlisting}[ language=SQL,
					deletekeywords={IDENTITY},
					deletekeywords={[2]INT},
					morekeywords={CLUSTERED, SKYLINE, OF},
					framesep=8pt,
					xleftmargin=40pt,
					framexleftmargin=40pt,
					frame=tb,
					framerule=0pt ]
	SELECT * FROM Pokémon
	SKYLINE OF Rareté MIN, Durée MIN, Victoire MAX
\end{lstlisting}

Et, la requête associée sans opérateur Skyline est la suivante~:
\lstset{style=SQLStyle}
\begin{lstlisting}[ language=SQL,
					deletekeywords={IDENTITY},
					deletekeywords={[2]INT},
					morekeywords={CLUSTERED, SKYLINE, OF},
					framesep=8pt,
					xleftmargin=40pt,
					framexleftmargin=40pt,
					frame=tb,
					framerule=0pt ]
    SELECT * FROM Pokémon AS P1
    WHERE NOT EXISTS (
        SELECT * FROM Pokémon AS P2
        WHERE (P2.Rareté <= P1.Rareté
          AND P2.Durée <= P1.Durée
          AND P2.Victoire >= P1.Victoire)
          AND (P2.Rareté < P1.Rareté
           OR P2.Durée < P1.Durée
           OR P2.Victoire > P1.Victoire));
\end{lstlisting}

Dans une représentation graphique, spécifiée par la dominance de Pareto, nous appelons point appartenant à un Skyline, ou point du Skyline, ou encore point Pareto-optimal, $sp$, chacun de ces tuples, et leur ensemble forme un Skyline $S$.

\begin{example}
	Avec notre exemple, l'ensemble obtenu est composé des tuples $(1, 5, 20, 70)$, $(2, 4, 60, 50)$ et $(4, 1, 80, 60)$. Il s'agit de l'ensemble des séquences de Pokémon dans un combat qui sont aussi bons ou meilleurs selon toutes les dimensions critères considérées (\texttt{Rareté}, \texttt{Durée} et \texttt{Victoire}) et meilleur pour au moins l'un de ces critères.
\end{example}

\section{Classement de Skyline}

De nombreux travaux ont été consacrés à l'étude du classement de Skyline. Dans \cite{chanHighDimensionalSkylines2006}, une métrique appelée fréquence Skyline est proposée afin d'ordonner un Skyline en fonction des points avec une haute fréquence Skyline. Cette méthode s'adapte bien à un grand nombre de dimensions et les expérimentations présentent une belle efficacité de l'algorithme.

Les requêtes top-$k$ (\cite{yiuEfficientProcessingTopk2007})\footnote{top-$k$ est une méthode de classement avec fonction d'évaluation.} peuvent servir d'alternative aux requêtes Skyline. Une autre technique se base sur la définition d'une «~forme~» représentant la recherche de l'utilisateur en spécifiant des régions définies par le décideur qui dominent toutes les autres régions (\cite{bartoliniFlexibleIntegrationMultimedia2007}). Une approche de classement de Skyline pour un Skycube (\cite{lakhalMultidimensionalSkylineAnalysis2017}) se concentrant sur les points de Skyline les plus chargés en information a aussi été proposée (\cite{vlachouRankingSkyDiscovering2010}). Cette méthode capture les relations de dominance entre les points de Skyline appartenant à différents sous-espaces. Un nouvel opérateur a aussi été introduit afin de trouver le point de Skyline le plus avantageux (\cite{gaoEfficientAlgorithmsFinding2015}).

\subsection{Méthode dp-idp}\index{Dp-idp}

dp-idp (pour \emph{dominance power and inverse dominance power}) est inspirée du schéma de pondération tf-idf (pour \emph{term frequency-inverse document frequency}) utilisé en recherche d'information, qui attribue à un terme $t$ un poids dans un document $d$. L'idée sous-jacente n'est pas de déterminer le nombre d'occurrences de chaque terme de la requête $t$ dans $d$, mais plutôt le poids tf-idf de chaque terme dans $d$. L'objectif étant de trouver des mots-clefs importants dans un corpus documentaire. Dans le contexte Skyline, les points dominés ont des impacts différents sur les points du Skyline. Ainsi, leur contribution dépend de caractéristiques locales correspondant à des points du Skyline et de caractéristiques globales correspondant au Skyline entier. La dominance d'un point est inversement proportionnelle au nombre de points qui le domine (\cite{valkanasSkylineRankingIR2014}), \emph{i.e.}~:
\begin{equation}\label{eq:dp}
	dp(p, sp) = \frac{1}{lm(p,sp)}
\end{equation}

dp-idp prend en compte les positions relatives des points dominés pour les différencier en se concentrant sur les points qui sont peu dominés~: \emph{e.g.} soit $sp$ un point d'un Skyline, si $sp \prec_d p_1, sp \prec_d p_2$ et ni $p_1$ ni $p_2$ ne dominent l'autre, ils sont similaires par rapport à $sp$. Sinon, si $p_1 \prec_d p_2$, alors $score(p_1) > score(p_2)$, et par conséquent la contribution de $p_1$ est plus importante. L'$idp$ d'un point $p \in r \backslash S$ correspond au nombre de points du Skyline qui dominent $p$. Moins un point $p$ apparaît de manière fréquente dans un ensemble de points dominés d'un Skyline, plus il est considérable~:
\begin{equation}\label{eq:idp}
	idp(p) = \log \frac{|S|}{|\{sp \in S : sp \prec_d p\}|}
\end{equation}

Afin de calculer la valeur $dp$ d'un point dominé $p$, sa position relative par rapport à un point du Skyline $sp$ est essentielle. Ainsi, un même point dominé peut contribuer différemment à différents points du Skyline. Ainsi, il est nécessaire de calculer la couche de \emph{minima} (ou \emph{layer of minima})\footnote{Couche de \emph{minima} est un concept analogue à celui de couche de \emph{maxima}, plus commun. Il s'agit de tous les points minimaux d'un ensemble donné. C'est la première couche, ou «~frontière~», de points qui ne domine aucun autre de l'espace multidimensionnel.} $lm(p, sp)$ où se situe le point dominé $p$ par rapport à $sp$. Le pouvoir de dominance de $p$ est alors l'inverse de la valeur de sa «~couche~»~. Le $Score(sp)$ qui mesure l'importance d'un point d'un Skyline $sp$ est défini de la manière suivante~:
\begin{equation}\label{eq:score}
	Score(sp) = \sum_{p : sp \prec_d p} dp(p, sp) \cdot idp(p)
\end{equation}

Les étapes de l'approche naïve permettant de classer un Skyline grâce à dp-idp sont~:
\begin{itemize}
	\item calcul des couches de \emph{minima} des points du Skyline $sp$~;
	\item mise à jour de $Score(sp)$ en considérant pour chaque point $p$ dans chaque couche de \emph{minima} $lm(p, sp)$ le nombre de points qui le dominent~;
	\item ordonner le Skyline en fonction des différents calculs.
\end{itemize}

Malheureusement, cette approche est peu efficace, et elle est notamment assez couteuse en temps en raison de calculs répétés. Elle est également dépourvue de toute notion de progressivité puisque nous devons d'abord ordonner le Skyline entier (\cite{valkanasSkylineRankingIR2014}).

Pour ces raisons, une approche alternative plus efficace, SkyIR, a été proposée (\cite{valkanasSkylineRankingIR2014}). L'algorithme a été décliné en fonction du modèle de priorité utilisé (\emph{Round-robin}, priorité par nombre des points non encore traités ou priorité par borne supérieure, ou \emph{Upper Bound (UBS)}). La configuration la plus avantageuse est presque toujours celle avec le système de priorité \emph{UBS}, soit SkyIR-UBS.

Le principal défaut de SkyIR-UBS apparaît être la complexité induite par les calculs de toutes les couches de \emph{minima} qu'il effectue. Nous proposons de perfectionner cette approche en ne prenant en compte, pour chaque point du Skyline, que ses dominés les plus proches, et plus aucun de ses dominés «~indirects~». De la sorte, les $lm(p, sp)$ sont composés d'un $sp$ et de la suite des points directement dominés de $sp$ jusqu'à $p$.

\subsection{Amélioration de dp-idp}

La relation de dominance peut être vue comme un tri hiérarchique. Autrement dit, un point d'un Skyline a nécessairement une position hiérarchique supérieure à celles des points qu'il domine. Cela nous a encouragé à mettre en correspondance la relation de dominance vue ci-dessus avec un graphe que nous appelons hiérarchie de dominance. L'utilisation d'une hiérarchie de dominance au sein de la méthode de classification dp-idp permet un calcul bien plus rapide des couches de \emph{minima}, le graphe étant élagué de ses arêtes inutiles lors de son parcours, et conduit, par conséquent, à une plus grande efficacité. Le graphe que nous proposons est un graphe orienté acyclique donnant une représentation d'un ensemble partiellement ordonné en établissant son graphe de couverture\footnote{Un graphe de couverture d'un graphe est un graphe couvrant tous les sommets du graphe d'origine en utilisant pour cela le moins d'arêtes possible.}. Un graphe orienté acyclique offre un ordre topologique qui peut donner une excellente représentation d'une hiérarchie d'un point de Skyline $sp$ par rapport aux points qu'il domine.

\begin{definition}[Hiérarchie de dominance]\index{Hiérarchie de dominance}
	Soit un ensemble de points $D$ et un ordre de dominance $\prec_d$, alors la hiérarchie de dominance (HD, ou DH pour \emph{dominance hierarchy}) est le graphe de couverture de l'ensemble ordonné $(D, \prec_d)$.
\end{definition}

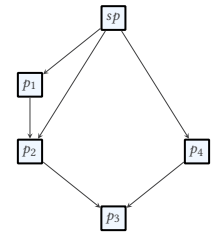
\begin{figure}[htb]
	\centering
	\resizebox{.15\textwidth}{!}{
		\begin{tikzpicture}[
			line join=bevel,
			box/.style={rectangle, draw=black, fill=aliceblue, very thick, minimum size=5mm}
			]
			
			\node [box] (p3) at (150pt, 0pt) {$p_3$};
			\node [box] (p2) at (100pt, 40pt) {$p_2$};
			\node [box] (p4) at (200pt, 40pt) {$p_4$};
			\node [box] (p1) at (100pt, 80pt) {$p_1$};
			\node [box] (sp) at (150pt, 120pt) {$sp$};
			
			\draw [stealth-] (p3) -- (p2);
			\draw [stealth-] (p3) -- (p4);
			\draw [stealth-] (p2) -- (p1);
			\draw [stealth-] (p2) -- (sp);
			\draw [stealth-] (p4) -- (sp);
			\draw [stealth-] (p1) -- (sp);
		\end{tikzpicture}
	}
	\caption{Exemple de graphe de hiérarchie de dominance}\label{fig:exemple_de_graphe_de_hierarchie_de_dominance}
\end{figure}

\begin{example}
	La figure~\ref{fig:exemple_de_graphe_de_hierarchie_de_dominance} représente une hiérarchie de dominance ayant comme sommet le point du Skyline $sp$. L'ordre de dominance est illustré par les arrêtes entre $sp$ et les points qu'il domine ($p_1, p_2, p_3$ et $p_4$).
\end{example}

Nous considérons la couche de \emph{minima} $lm(p, sp)$ comme le nombre de sommets du chemin minimal entre $sp$ et $p$ dans la hiérarchie de dominance.

\begin{example}
	Pour calculer la couche \emph{minima} $lm(p3, sp)$, nous voyons sur la figure~\ref{fig:exemple_de_graphe_de_hierarchie_de_dominance} qu'il y a deux chemins de $sp$ jusqu'à $p_3$~:
	\begin{enumerate}
		\item Premier chemin~: $sp \to p_1 \to p_2 \to p_3$.
		\item Second chemin~: $sp \to p_4 \to p_3$.
	\end{enumerate}	
	
	Le premier chemin passe par quatre sommets alors que le second n'en compte que trois, donc le chemin minimal de $sp$ à $p_3$ est le second et $lm(p_3, sp) = 3$.
\end{example}

\begin{figure}[htb]
	\centering
	\resizebox{.3\textwidth}{!}{
		\begin{tikzpicture}[
			line join=bevel,
			box/.style={rectangle, draw=black, fill=aliceblue, very thick, minimum size=5mm}
			]
			
			\node [box] (p01) at (150pt, 0pt) {$8$};
			\node [box] (p11) at (50pt, 40pt) {$5$};
			\node [box] (p12) at (150pt, 40pt) {$7$};
			\node [box] (p13) at (250pt, 40pt) {$6$};
			\node [box] (p21) at (250pt, 80pt) {$3$};
			\node [box] (p31) at (50pt, 120pt) {$4$};
			\node [box] (p32) at (150pt, 120pt) {$2$};
			\node [box] (p33) at (250pt, 120pt) {$1$};
			\node [box] (i) at (150pt, 160pt) {$I+$};
			
			\draw [stealth-] (p01) -- (p11);
			\draw [stealth-] (p01) -- (p12);
			\draw [stealth-] (p01) -- (p13);
			\draw [stealth-] (p11) -- (p31);
			\draw [stealth-] (p11) -- (p32);
			\draw [stealth-] (p11) -- (p21);
			\draw [stealth-] (p12) -- (p31);
			\draw [stealth-] (p12) -- (p32);
			\draw [stealth-] (p12) -- (p21);
			\draw [stealth-] (p13) -- (p21);
			\draw [stealth-] (p21) -- (p33);
			\draw [stealth-] (p31) -- (i);
			\draw [stealth-] (p32) -- (i);
			\draw [stealth-] (p33) -- (i);
		\end{tikzpicture}
	}
	\caption{Graphe de hiérarchie de dominance de l'exemple}
	\label{fig:graphe_de_hierarchie_de_dominance_de_pokemon_showdown}
\end{figure}
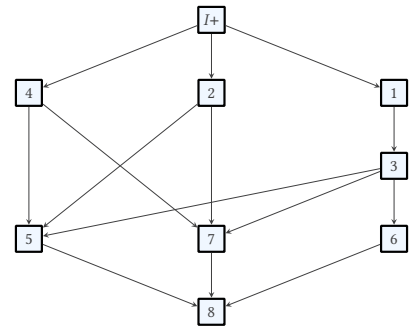

\begin{example}
	En considérant à nouveau la relation \texttt{Pokémon} (cf. tableau~\ref{tab:relation_exemple}), le graphe de hiérarchie de dominance illustrant les relations de dominance entre les points du Skyline et les points dominés est donné par la figure~\ref{fig:graphe_de_hierarchie_de_dominance_de_pokemon_showdown} ($I^+$ étant le point idéal théorique, ou abstrait, qui domine tous les points du Skyline, et les points sont représentés par leur \texttt{RowId}).
	
	En considérant à nouveau la relation \texttt{Pokémon} (cf. tableau~\ref{tab:relation_exemple}), pour calculer le score de chaque point du Skyline, nous utilisons la formule~\ref{eq:score}. Ainsi, le classement par ordre décroissant des points du Skyline, représentés par leur \texttt{RowId}, est soit $1, 2, 4$, soit $1, 4, 2$.
	
	La formule de calcul des idp a pour particularité que chaque point dominés par tous les points du Skyline aura un score de $0$. En effet, ces points ne modifient pas le classement du Skyline car ils affectent tous les points du Skyline de la même manière (\cite{valkanasSkylineRankingIR2014}).
	
	Sur la base des résultats obtenus (cf. tableau~\ref{tab:calcul_de_score_avec_la_methode_dp_idp_amelioree}), nous montrons que la méthode dp-idp n'offre pas toujours la possibilité de distinguer deux points du Skyline comme c'est le cas ici concernant ceux de valeur de \texttt{RowId} $2$ et $4$.
	
	\begin{table}[htb]
		\caption{Calcul de score avec la méthode dp-idp améliorée}\label{tab:calcul_de_score_avec_la_methode_dp_idp_amelioree}
		\centering
		\begin{minipage}{\linewidth}
			\centering
			\small
			\begin{tabular}{c|c|c|c}
				\toprule
				\texttt{RowId} des $sp$ & \texttt{RowId} des $p$ & $lm(p, sp)$ & $Score(sp)$ \\
				\midrule
				$1$ & $3$ & $lm(3, 1) = 2$ & $0.398$ \\
				& $5$ & $lm(5, 1) = 3$ &         \\
				& $6$ & $lm(6, 1) = 3$ &         \\
				& $7$ & $lm(7, 1) = 3$ &         \\
				& $8$ & $lm(8, 1) = 4$ &         \\
				\midrule
				$2$ & $5$ & $lm(5, 2) = 2$ & $0$     \\
				& $7$ & $lm(7, 2) = 2$ &         \\
				& $8$ & $lm(8, 2) = 3$ &         \\
				\midrule
				$4$ & $5$ & $lm(5, 4) = 2$ & $0$     \\
				& $7$ & $lm(7, 4) = 2$ &         \\
				& $8$ & $lm(8, 4) = 3$ &         \\
				\bottomrule
			\end{tabular}
		\end{minipage}
	\end{table}
\end{example}

\subsubsection{Algorithme de la méthode améliorée}

L'algorithme dp-idp avec hiérarchie de dominance est un algorithme qui fait appel à quatre sous-algorithmes~: matriceDesDominants, grapheDeCouverture, lm et score$_{dp-idp}$. Il est décrit dans l'algorithme~\ref{algo:dp-idp}.

\begin{algorithm}[htbp]
	\caption{Algorithme dp-idp avec hiérarchie de dominance\label{algo:dp-idp}}
	\begin{algorithmic}
		\INPUT~ \\
		La relation $r$. \\
		\OUTPUT~ \\
		Le tableau des scores de dp-idp $score$. \\
		\STATE
		\COMMENT{Appel au quatre sous-algorithmes}
		\STATE $m_{\prec_d}, S_{\setminus \mathcal{D}+lm} := \texttt{matriceDesDominants}(r)$;
		\STATE $m_{\prec_d}, S_{\setminus \mathcal{D}+lm}, S_{\prec_d}C, idpC := $
		\STATE $\texttt{grapheDeCouverture}(m_{\prec_d}, S_{\setminus \mathcal{D}+lm})$;
		\STATE $S_{\setminus \mathcal{D}+lm} := \texttt{lm}(m_{\prec_d}, S_{\setminus \mathcal{D}+lm}, S_{\prec_d}C)$;
		\RETURN $\texttt{score}_{dp-idp}(S_{\setminus \mathcal{D}+lm}, idpC)$;
	\end{algorithmic}
\end{algorithm}

\begin{example}
	Le tableau~\ref{tab:relation_exemple_2} représente la relation \texttt{Pokémon} qui, pour des raisons de commodité, ne conserve aucun attribut ou commentaire, seulement le \texttt{RowId} des tuples.
	
	\begin{table}[htb]
		\caption{La relation $\texttt{Pokémon}_{1}$ simplifiée}\label{tab:relation_exemple_2}
		\centering
		\begin{minipage}{\linewidth}
			\centering
			\small
			\begin{tabular}{c|ccc}
				\toprule
				\texttt{RowId} & \texttt{Rareté} & \texttt{Durée} & \texttt{Victoire} \\
				\midrule
				$1$ & $5$ & $20$ & $70$ \\
				$2$ & $4$ & $60$ & $50$ \\
				$3$ & $5$ & $30$ & $60$ \\
				$4$ & $1$ & $80$ & $60$ \\
				$5$ & $5$ & $90$ & $40$ \\
				$6$ & $9$ & $30$ & $50$ \\
				$7$ & $7$ & $80$ & $40$ \\
				$8$ & $9$ & $90$ & $30$ \\
				\bottomrule
			\end{tabular}
		\end{minipage}
	\end{table}
\end{example}

Le sous-algorithme matriceDesDominants, décrit dans l'algorithme~\ref{algo:matrice_des_dominants}, génère, depuis la relation $r$, un tableau à deux dimensions carré, ou matrice, $m_{\prec_d}$, de la cardinalité de $r$. $m_{\prec_d}$ indique les dominances, \emph{i.e.} en colonne est signalé le \texttt{RowId} du tuple dominant le tuple représenté en ligne par son \texttt{RowId}. Le sous-algorithme est donné dans le cas où toutes les préférences Skyline sont \texttt{MIN}, mais il est aisément transposable pour n'importe quelle combinaisons de préférences Skyline. Le Skyline $S_{\setminus \mathcal{D}+lm}$ est aussi calculé à cette occasion, et il est retourné sans dimension et destiné à recevoir les couches \emph{minima}.

\begin{algorithm}[htbp]
	\caption{Algorithme matriceDesDominants ($\mathcal{O}(|r|^2 \cdot |\mathcal{D}|)$)\label{algo:matrice_des_dominants}}
	\begin{algorithmic}
		\INPUT~ \\
		La relation $r$. \\
		\OUTPUT~ \\
		Le tableau à deux dimensions carré indiquant les dominances $m_{\prec_d}$. \\
		Le Skyline sans dimension destiné à recevoir les $lm(sp, p)$ des points dominés $S_{\setminus \mathcal{D}+lm}$. \\
		\LET $\mathcal{D} := \{d_1, \dotsc, d_n \}$~: l'ensemble des dimensions de $r$.
		\LET $m_{\prec_d}$~: un tableau de $|r|$ tableaux de $|r|$ booléens faux
		\LET $idpC$~: une relation sans dimension de $|r|$ tuples avec les mêmes $\texttt{RowId}$ que ceux de $r$
		\FOR{$i := 0, \dots, |r| - 1$}
		\FOR{$j := 0, \dots, |r| - 1$}
		\IF{$i \neq j$}
		\LET $t_i$~: $t_i \in r, t_i[\texttt{RowId}] = i$;
		\LET $t_j$~: $t_j \in r, t_j[\texttt{RowId}] = j$;
		\LET $sup := vrai$;
		\FORALL{$d_k \in \mathcal{D}$}
		\STATE
		\COMMENT{Calcul de dominance}
		\STATE
		\COMMENT{Comparateur $>$ si $\forall \{d_1, \dotsc, d_n \} \in \mathcal{D}, Pref(d_k) = MIN$}
		\IF{$t_j[d_k] > t_i[d_k]$}
		\STATE $sup := faux$;
		\COMMENT{$t_j \nprec_d t_i$}
		\EXITFOR
		\ENDIF
		\ENDFOR
		\IF{$sup$}
		\LET $t'_i$~: $t'_i \in S_{\setminus \mathcal{D}}, t'_i[\texttt{RowId}] = i$;
		\STATE
		\COMMENT{Ajout de l'arête au graphe}
		\STATE $m_{\prec_d}[i][j] := vrai$;
		\COMMENT{$t_j \prec_d t_i$}
		\STATE $S_{\setminus \mathcal{D}} := S_{\setminus \mathcal{D}} \setminus t'_i$;
		\COMMENT {$t_i \notin S$}
		\ENDIF
		\ENDIF
		\ENDFOR
		\ENDFOR
		\LET $S_{\setminus \mathcal{D}+lm}$~: une relation de $|r|$ dimensions de $S_{\setminus \mathcal{D}+lm}$ tuples $(0, \dots, 0)$ avec les mêmes $\texttt{RowId}$ que ceux de $S_{\setminus \mathcal{D}+lm}$
		\RETURN $m_{\prec_d}, S_{\setminus \mathcal{D}+lm}$;
	\end{algorithmic}
\end{algorithm}

\begin{example}
	En considérant à nouveau la relation \texttt{Pokémon} (cf. tableau~\ref{tab:relation_exemple_2}), résultant de l'algorithme~\ref{algo:matrice_des_dominants}, le tableau~\ref{tab:matrice_des_dominants} montre la matrice des dominants $m_{\prec_d}$ et le tableau~\ref{tab:skyline_sans_dimension_initial} montre le Skyline, sans dimension et destiné à recevoir les couches \emph{minima} $S_{\setminus \mathcal{D}+lm}$. Avec le tableau~\ref{tab:matrice_des_dominants}, nous voyons que le tuple de \texttt{RowId} $6$ est dominé à la fois par celui de de \texttt{RowId} $1$ et celui de \texttt{RowId} $3$. Le tableau~\ref{tab:skyline_sans_dimension_initial} nous confirme que le Skyline est composé des tuples de  \texttt{RowId} $1$, $2$ et $4$, visibles en ligne, alors que les \texttt{RowId} de l'ensemble de tuple de la relation sont donnés en colonnes.
	
	\begin{table}[htb]
		\caption{La matrice des dominants $m_{\prec_d}$}\label{tab:matrice_des_dominants}
		\centering
		\begin{minipage}{\linewidth}
			\centering
			\small
			\begin{tabular}{c|c|c|c|c|c|c|c|c}
				\toprule
				\texttt{RowId} & \texttt{$1$} & \texttt{$2$} & \texttt{$3$} & \texttt{$4$} & \texttt{$5$} & \texttt{$6$} & \texttt{$7$} & \texttt{$8$} \\
				\midrule
				$1$ &  &  &  &  &  &  &  &  \\
				\midrule
				$2$ &  &  &  &  &  &  &  &  \\
				\midrule
				$3$ & \checkmark &  &  &  &  &  &  &  \\
				\midrule
				$4$ &  &  &  &  &  &  &  &  \\
				\midrule
				$5$ & \checkmark & \checkmark & \checkmark & \checkmark &  &  &  &  \\
				\midrule
				$6$ & \checkmark &  & \checkmark &  &  &  &  &  \\
				\midrule
				$7$ & \checkmark & \checkmark & \checkmark & \checkmark &  &  &  &  \\
				\midrule
				$8$ & \checkmark & \checkmark & \checkmark & \checkmark & \checkmark & \checkmark & \checkmark &  \\
				\bottomrule
			\end{tabular}
		\end{minipage}
	\end{table}

	\begin{table}[htb]
		\caption{Le Skyline sans dimension voué à avoir les \emph{lm} $S_{\setminus \mathcal{D}+lm}$}\label{tab:skyline_sans_dimension_initial}
		\centering
		\begin{minipage}{\linewidth}
			\centering
			\small
			\begin{tabular}{c|cccccccc}
				\toprule
				\texttt{RowId} & \texttt{$1$} & \texttt{$2$} & \texttt{$3$} & \texttt{$4$} & \texttt{$5$} & \texttt{$6$} & \texttt{$7$} & \texttt{$8$} \\
				\midrule
				$1$ & 0 & 0 & 0 & 0 & 0 & 0 & 0 & 0 \\
				$2$ & 0 & 0 & 0 & 0 & 0 & 0 & 0 & 0 \\
				$4$ & 0 & 0 & 0 & 0 & 0 & 0 & 0 & 0 \\
				\bottomrule
			\end{tabular}
		\end{minipage}
	\end{table}
\end{example}

Le sous-algorithme grapheDeCouverture, décrit dans l'algorithme~\ref{algo:graphe_de_couverture}, a pour objectif de mettre à jour la matrice des dominants $m_{\prec_d}$ en ne considérant plus que le graphe de couverture. Le Skyline sans dimension $S_{\setminus \mathcal{D}+lm}$ est aussi mis à jour afin d'être prêt à recevoir les couches \emph{minima}, chacune des couches à calculer étant indiquée par la valeur $1$. A des fins d'optimisation, les cardinalités de dominance des points du Skyline $S_{\prec_d}C$ ainsi que les cardinalités utiles à la mesure des idp, $idpC$, sont également calculés durant le parcours du graphe de couverture.

\begin{algorithm}[htbp]
	\caption{Algorithme grapheDeCouverture ($\mathcal{O}(|r|^3/2)$)\label{algo:graphe_de_couverture}}
	\begin{algorithmic}
		\INPUT~ \\
		Le tableau à deux dimensions carré indiquant les dominances $m_{\prec_d}$. \\
		Le Skyline sans dimension destiné à recevoir les $lm(sp, p)$ des points dominés $S_{\setminus \mathcal{D}+lm}$. \\
		\OUTPUT~ \\
		Le tableau à deux dimensions carré indiquant les dominances du graphe de couverture $m_{\prec_d}$. \\
		Le Skyline sans dimension prêt à recevoir les $lm(sp, p)$ des points dominés $S_{\setminus \mathcal{D}+lm}$. \\
		Les cardinalités de dominance des points de $S$ $S_{\prec_d}C$. \\
		Les cardinalités des $sp \in S, sp \prec_d p$ $idpC$. \\
		\LET $S_{\prec_d}C$~: un tableau de $|S_{\setminus \mathcal{D}+lm}|$ entiers 0
		\LET $idpC$~: un tableau de $|m_{\prec_d}|$ entiers 0
		\FOR{$i := 0, \dots, |m_{\prec_d}| - 1$}
		\IF{$t_i, t[\texttt{RowId}] = i, t_i \notin S_{\setminus \mathcal{D}+lm}$}
		\FOR{$j := 0, \dots, |m_{\prec_d}| - 1$}
		\IF{$m_{\prec_d}[i][j]$}
		\IF{$t_j, t[\texttt{RowId}] = j, t_j \in S_{\setminus \mathcal{D}+lm}$}
		\STATE
		\COMMENT{Marquage des dominances}
		\STATE $t_j[i] := 1$
		\STATE
		\COMMENT{Mise à jour des cardinalités de dominance}
		\STATE $S_{\prec_d}C[j] := S_{\prec_d}C[j] + 1$;
		\STATE $idpC[i] := idpC[i] + 1$;
		\ELSE
		\FOR{$k = 0, \dots, |m_{\prec_d}| - 1$}
		\IF{$m_{\prec_d}[j][k]$}
		\STATE
		\COMMENT{Suppression de l'arête inutile du graphe de couverture}
		\STATE $m_{\prec_d}[j][k] := faux$;
		\ENDIF
		\ENDFOR
		\ENDIF
		\ENDIF
		\ENDFOR
		\ENDIF
		\ENDFOR
		\RETURN $m_{\prec_d}, S_{\setminus \mathcal{D}+lm}, S_{\prec_d}C, idpC$;
	\end{algorithmic}
\end{algorithm}

\begin{example}
	En considérant à nouveau la relation \texttt{Pokémon} (cf. tableau~\ref{tab:relation_exemple_2}), résultant de l'algorithme~\ref{tab:matrice_des_dominants_du_graphe_de_couverture}, le tableau~\ref{tab:matrice_des_dominants_du_graphe_de_couverture} montre la matrice des dominants $m_{\prec_d}$ selon le graphe de couverture et le tableau~\ref{tab:skyline_sans_dimension_pret_pour_lm} montre le Skyline, sans dimension et prêt à recevoir les couches \emph{minima} $S_{\setminus \mathcal{D}+lm}$, fusionné, en colonne, avec les cardinalités de dominance des points du Skyline $S_{\prec_d}C$ et, en ligne, avec les cardinalités utiles à la mesure des idp, $idpC$. Cette présentation nous apparaît plus pratique et a pour objectif d'améliorer aussi la compréhension. Avec le tableau~\ref{tab:matrice_des_dominants_du_graphe_de_couverture} nous retrouvons le graphe de hiérarchie de dominance de Pokémon Showdown! (cf. figure~\ref{fig:graphe_de_hierarchie_de_dominance_de_pokemon_showdown}). Et, avec le tableau~\ref{tab:skyline_sans_dimension_pret_pour_lm}, nous voyons (avec les valeurs $1$) toutes les dominances de chaque point du Skyline (dont les \texttt{RowId} sont en ligne). De même, nous voyons le nombre de points dominés par chaque point du Skyline dans la colonne $S_{\prec_d}C$, et le nombre de points du Skyline dominant chaque point de la relation (dont les \texttt{RowId} sont en colonne) avec la ligne unique du tableau $idpC$.
	
	\begin{table}[htb]
		\caption{$m_{\prec_d}$ d'après le graphe de couverture}\label{tab:matrice_des_dominants_du_graphe_de_couverture}
		\centering
		\begin{minipage}{\linewidth}
			\centering
			\small
			\begin{tabular}{c|c|c|c|c|c|c|c|c}
				\toprule
				\texttt{RowId} & \texttt{$1$} & \texttt{$2$} & \texttt{$3$} & \texttt{$4$} & \texttt{$5$} & \texttt{$6$} & \texttt{$7$} & \texttt{$8$} \\
				\midrule
				$1$ &  &  &  &  &  &  &  &  \\
				\midrule
				$2$ &  &  &  &  &  &  &  &  \\
				\midrule
				$3$ & \checkmark &  &  &  &  &  &  &  \\
				\midrule
				$4$ &  &  &  &  &  &  &  &  \\
				\midrule
				$5$ &  & \checkmark & \checkmark & \checkmark &  &  &  &  \\
				\midrule
				$6$ &  &  & \checkmark &  &  &  &  &  \\
				\midrule
				$7$ &  & \checkmark & \checkmark & \checkmark &  &  &  &  \\
				\midrule
				$8$ &  &  &  &  & \checkmark & \checkmark & \checkmark &  \\
				\bottomrule
			\end{tabular}
		\end{minipage}
	\end{table}
	
	\begin{table}[htb]
		\caption{$S_{\setminus \mathcal{D}+lm}$ prêt à recevoir les \emph{lm} avec $S_{\prec_d}C$ et $idpC$ }\label{tab:skyline_sans_dimension_pret_pour_lm}
		\centering
		\begin{minipage}{\linewidth}
			\centering
			\small
			\begin{tabular}{c|cccccccc|c}
				\toprule
				\texttt{RowId} & \texttt{$1$} & \texttt{$2$} & \texttt{$3$} & \texttt{$4$} & \texttt{$5$} & \texttt{$6$} & \texttt{$7$} & \texttt{$8$} & $S_{\prec_d}C$ \\
				\midrule
				$1$    & 0 & 0 & 1 & 0 & 1 & 1 & 1 & 1 & 5 \\
				$2$    & 0 & 0 & 0 & 0 & 1 & 0 & 1 & 1 & 3 \\
				$4$    & 0 & 0 & 0 & 0 & 1 & 0 & 1 & 1 & 3 \\
				\midrule
				\multicolumn{1}{c|}{$idpC$} & 
				\multicolumn{1}{c|}{0} & 
				\multicolumn{1}{c|}{0} & 
				\multicolumn{1}{c|}{1} & 
				\multicolumn{1}{c|}{0} & 
				\multicolumn{1}{c|}{3} & 
				\multicolumn{1}{c|}{1} & 
				\multicolumn{1}{c|}{3} & 
				\multicolumn{1}{c|}{3} & / \\
				\bottomrule
			\end{tabular}
		\end{minipage}
	\end{table}
\end{example}

Le sous-algorithme lm, décrit dans l'algorithme~\ref{algo:lm}, calcule les couches \emph{minima} pour chaque point du Skyline selon le graphe de couverture indiqué dans la matrice des dominants $m_{\prec_d}$. Les résultats sont enregistrés dans le Skyline sans dimension $S_{\setminus \mathcal{D}+lm}$. L'utilisation d'une fonction récursive serait toute indiquée pour ce sous-algorithme, mais son usage le rendrait moins efficace. Aussi, nous préférons ne présenter que la version sans récursivité.

\begin{algorithm}[htbp]
	\caption{Algorithme lm ($\mathcal{O}(|r| \cdot |S|)$)\label{algo:lm}}
	\begin{algorithmic}
		\INPUT~ \\
		Le tableau à deux dimensions carré indiquant les dominances $m_{\prec_d}$. \\
		Le Skyline sans dimension prêt à recevoir les $lm(sp, p)$ des points dominés $S_{\setminus \mathcal{D}+lm}$. \\
		Les cardinalités de dominance des points de $S$ $S_{\prec_d}C$. \\
		\OUTPUT~ \\
		Le Skyline $S$ avec les $lm(sp, p)$ des points dominés $S_{lm}$. \\
		
		
		\FORALL{$t \in S_{\setminus \mathcal{D}+lm}$}
		\LET $i := t[\texttt{RowId}]$;
		\STATE $couche := i$
		\STATE $prof := 2$
		\WHILE{$S_{\prec_d}C[i] > 0$}
		\STATE $couche_{+1} := couche$
		\FOR{$j = 0, \dots, |m_{\prec_d}| - 1$}
		\IF{$m_{\prec_d}[j][couche] \wedge t[j] > 0$}
		\IF{$t[j] = 1 \vee t[j] \neq 1 \wedge t[j] > prof$}
		\STATE
		\COMMENT{Ajustement de la profondeur de $lm(t, t_j), t_j \in r$}
		\STATE $t[j] := t[j] + prof$;
		\ENDIF
		\STATE $S_{\prec_d}C[i] := S_{\prec_d}C[i] - 1$;
		\COMMENT{Dominance traitée}
		\STATE $couche_{+1} := j$
		\IF{$S_{\prec_d}C[i] = 0$}
		\STATE
		\COMMENT{Le traitement du point du Skyline s'arrête lorsque toutes ses dominances sont traitées}
		\EXITFOR
		\ENDIF
		\ENDIF
		\STATE $prof := prof + 1$
		\IF{$couche_{+1} = couche$}
		\EXITFOR
		\COMMENT{Il n'y a plus de couche à traiter}
		\ENDIF
		\STATE $couche := couche_{+1}$
		\ENDFOR
		\ENDWHILE
		\ENDFOR
		\RETURN{$S_{\setminus \mathcal{D}+lm}$};
	\end{algorithmic}
\end{algorithm}

\begin{example}
	En considérant à nouveau la relation \texttt{Pokémon} (cf. tableau~\ref{tab:relation_exemple_2}), résultant de l'algorithme~\ref{algo:lm}, le tableau~\ref{tab:skyline_sans_dimension_final} montre le Skyline sans dimension mais avec les couches \emph{minima} $S_{\setminus \mathcal{D}+lm}$. Nous rappelons que les \texttt{RowId} des tuples du Skyline sont affichés en ligne, alors que tous les \texttt{RowId} de tous les tuples de la relation le sont en colonne. De la sorte, nous voyons que $lm(3, 1) = 2$, $lm(8, 1) = 4$ ou encore que $lm(2, 5) = 2$.
	
	\begin{table}[htb]
		\caption{Le Skyline sans dimension avec les \emph{lm} $S_{\setminus \mathcal{D}+lm}$}\label{tab:skyline_sans_dimension_final}
		\centering
		\begin{minipage}{\linewidth}
			\centering
			\small
			\begin{tabular}{c|cccccccc}
				\toprule
				\texttt{RowId} & \texttt{$1$} & \texttt{$2$} & \texttt{$3$} & \texttt{$4$} & \texttt{$5$} & \texttt{$6$} & \texttt{$7$} & \texttt{$8$} \\
				\midrule
				$1$ & 0 & 0 & 2 & 0 & 3 & 3 & 3 & 4 \\
				$2$ & 0 & 0 & 0 & 0 & 2 & 0 & 2 & 3 \\
				$4$ & 0 & 0 & 0 & 0 & 2 & 0 & 2 & 3 \\
				\bottomrule
			\end{tabular}
		\end{minipage}
	\end{table}
\end{example}

Le sous-algorithme score$_{dp-idp}$, décrit dans l'algorithme~\ref{algo:score}, calcule les scores des points du Skyline selon la méthode dp-idp. Pour cela, il a besoin des couches \emph{minima} enregistrés dans le Skyline sans dimension $S_{\setminus \mathcal{D}+lm}$, utiles pour le calcul de dp, et il a aussi besoin, pour chaque point de la relation, du nombre de points du Skyline qui le domine, $idpC$, utiles pour le calcul de idp.

\begin{algorithm}[htbp]
	\caption{Algorithme score$_{dp-idp}$ ($\mathcal{O}(|S| \cdot |\mathcal{D}|)$)\label{algo:score}}
	\begin{algorithmic}
		\INPUT~ \\
		Le Skyline sans dimension avec les $lm(sp, p)$ des points dominés $S_{\setminus \mathcal{D}+lm}$. \\
		Les cardinalités des $sp \in S, sp \prec_d p$ $idpC$. \\
		\OUTPUT~ \\
		Le tableau des scores de dp-idp $score$. \\
		\LET $score$~: un tableau associatif de $|S_{\setminus \mathcal{D}+lm}|$ entiers 0
		\FORALL{$t \in S_{\setminus \mathcal{D}+lm}$}
		\LET $i := t[\texttt{RowId}]$;
		\FOR{$j = 0, \dots, |t| - 1$}
		\IF{$t[j] > 0$}
		\STATE $score[i] := score[i] + 1 / t[j] \times \log(|S_{\setminus \mathcal{D}+lm}| / idpC[j])$;
		\STATE
		\COMMENT{cf. formule~\ref{eq:score}}
		\ENDIF
		\ENDFOR
		\ENDFOR
		\RETURN $score$;
	\end{algorithmic}
\end{algorithm}

\begin{example}
	En considérant à nouveau la relation \texttt{Pokémon} (cf. tableau~\ref{tab:relation_exemple_2}), résultant de l'algorithme~\ref{algo:score}, le tableau~\ref{tab:tableau_de_score_de_dp_idp_ameliore} montre les scores respectifs des tuples de \texttt{RowId} $1$, $2$ et $4$, qui compose le Skyline.
	
	\begin{table}[htbp]
		\caption{Le tableau des scores de dp-idp amélioré}\label{tab:tableau_de_score_de_dp_idp_ameliore}
		\centering
		\begin{minipage}{\linewidth}
			\centering
			\small
			\begin{tabular}{c|ccc}
				\toprule
				\texttt{RowId} & \texttt{$1$} & \texttt{$2$} & \texttt{$4$} \\
				\midrule
				$score$ & 0.398 & 0 & 0 \\
				\bottomrule
			\end{tabular}
		\end{minipage}
	\end{table}
\end{example}

\subsection{Méthode CoSky}\label{ssec:methode_CoSky}\index{CoSky}

Afin de proposer une solution de classement de Skyline qui permette de toujours différencier et d'ordonner des points dissociés d'un Skyline, nous présentons à présent la méthode CoSky. CoSky (pour cosinus Skyline) est une approche en plusieurs étapes qui n'utilise pas de relation de dominance ou de fonction mathématique gourmande en temps comme le logarithme\footnote{Le logarithme est une fonction transcendantale significativement plus couteuse, notamment en raison des optimisations matérielles fréquemment disponibles pour l'addition et la multiplication, que les opérations algébriques standards.}. C'est, à notre connaissance, la première méthode de type TOPSIS\footnote{TOPSIS pour \emph{technique for order preference by similarity to ideal solution}, est une méthode dont l'objectif est de classer par ordre de choix des alternatives sur la base de critères favorables ou défavorables.} (\cite{laiTOPSISMODM1994, behzadianStateofTheartSurvey2012}) appliqué à ce type de classement. TOPSIS est basé sur une normalisation vectorielle, un calcul de poids de chaque attribut, et un calcul de score de chaque point déterminé par une mesure géométrique des distances entre chaque alternative, représentée par un point, et les solutions idéales/anti-idéales. Dans la méthode CoSky, la normalisation des attributs est effectuée avec la somme, une pondération automatique des attributs normalisés selon l'indice de Gini, et le score utilise le cosinus de Salton de l'angle entre un point du Skyline et le point idéal.

Ce calcul est présenté en détails dans cette sous-section, et sa préparation ainsi que des remarques générales sont données en annexe. Pour chaque étape, nous considérons que $i \in [1..n]$ et $j \in [1..m]$ (où $n$ est le nombre de tuples et $m$ le nombre d'attributs).

\subsubsection{I. Normalisation par la somme}\label{sssec:etape_1_normalisation_des_attributs_par_la_somme}

Le Skyline est normalisé par la somme. Cette méthode garantit que toutes les valeurs normalisées sont comprises entre $-1$ et $1$, et que leur somme est égale à 1. Cela est utile lors de la représentation de données où la contribution relative de chaque valeur par rapport à l'ensemble est importante. Cette normalisation permet d'éliminer les anomalies liées à des unités de mesure et des échelles différentes tout en s'assurant que les attributs soient toujours mesurables et comparables entre eux.

Il est important de noter qu'il est préférable de ramener, préalablement, les valeurs non normalisées des attributs au dessus de $0$ afin d'éviter que les valeurs négatives soient confondues avec leurs opposées. En outre, de la sorte, chaque valeur d’attribut de chaque point du Skyline est convertie en une valeur sur une échelle allant de $0$ à $1$.

Soit $S_N$ l'ensemble des points de Skyline normalisés, ou Skyline normalisé, et le tuple $u_i = (u_{i}[A_1], u_{i}[A_2], \dotsc, u_{i}[A_m]) \in S_N$, alors nous avons~:
\begin{equation}
	u_{i}[A_j] = \frac{t_i [A_j]}{\sum_{i' = 1}^{n}t_{i'}[A_j]}, \forall t_i \in S
\end{equation} 

Cette méthode nécessite tout de même une vigilance afin d'éviter une division par $0$, sur la somme des valeurs non normalisées des attributs~: $\sum_{i' = 1}^{n}t_{i'}[A_j] \neq 0$.

\subsubsection{II. Pondération avec indice de Gini}\label{sssec:etape_2_ponderation_automatique_des_attributs_normalises_avec_l_indice_de_gini}

Le classement de Skyline vise le plus souvent à distinguer de manière stricte les points du Skyline. A cette fin, il est crucial de déterminer une mesure qui permette cette distinction. Pour ce faire, plusieurs mesures ont été proposées dans la littérature. Une méthode basée sur le concept d'entropie a ainsi été proposée dans des problèmes multicritères (\cite{jingwenhuangCombiningEntropyWeight2008, lotfiImpreciseShannonEntropy2010}). Cette méthode s'adapte bien à un contexte Skyline, pour lequel nous cherchons à différencier les valeurs des attributs par un classement de Skyline afin d'avoir une meilleure prise de décision. Cependant, elle a aussi ses limites, notamment en ce qui concerne le calcul de l'entropie qui nécessite l'usage d'une fonction logarithmique, gourmande en temps.

Nous préférons présenter une autre mesure, l'indice de Gini\footnote{L'indice (ou coefficient) de Gini est une mesure statistique utilisée pour évaluer l'inégalité d'une variable par rapport à une population donnée. Initialement, et toujours principalement, il est employé pour mesurer le degré d'inégalité des revenus d'un pays. Il varie entre 0 (égalité parfaite) et 1 (inégalité totale), et l'inégalité est d'autant plus forte que l'indice est élevé.}, qui est plus rapide à calculer que l'entropie et n'utilise pas de logarithme pour la pondération automatique des attributs. Dans la méthode présentée, l'indice de Gini est utilisé pour dériver les poids des attributs afin de déterminer le degré de divergence des valeurs des attributs. L'indice de Gini de $A_j$, $Gini(A_j)$, est déterminé à l'aide de l'équation suivante~:
\begin{equation}
	Gini(A_j) = 1 - \sum_{i = 1}^{n}u_{i}[A_j]^2
\end{equation}

Nous appelons $W$ le poids de l'attribut $A_j$. Le décideur peut spécifier directement $(W(A_1), W(A_2), \dotsc, W(A_m))$ de sorte que $\texttt{SUM}(W(A_1), W(A_2), \dotsc, W(A_m)) = 1$. Le poids d'un attribut est alors son importance, et est donné par la formule suivante~:
\begin{equation}
	W(A_j) = \frac{Gini(A_j)}{\sum_{j' = 1}^{m} Gini(A_{j'})} 
\end{equation}

Soit $S_P$ Skyline pondéré, ou l'ensemble des points de Skyline après pondération, et le tuple $v_i  = (v_{i}[A_1], v_{i}[A_2], \dotsc, v_{i}[A_m]) \in S_P$, alors nous avons~:
\begin{equation}
	v_{i}[A_j] = W (A_j) \times u_{i}[A_j], \forall u_i \in S_N
\end{equation}

\subsubsection{III. Détermination du point idéal}\label{sssec:etape_3_determination_du_point_ideal}

Le point idéal théorique, ou abstrait, noté $I^+$, qui domine tous les points du Skyline. Il correspond au tuple répondant de manière optimale aux préférences Skyline.

Ainsi, soit $I^+ = (I^+[A_1], I^+[A_2], \dotsc, I^+[A_m])$, alors nous avons~:
\begin{equation}
	I^+[A_j] = 
	\left\lbrace
	\begin{array}{l}
		\texttt{MAX}(v_i[A_j]) \equiv Pref(A_j) = \texttt{MAX} \\ 
		\texttt{MIN}(v_i[A_j]) \equiv Pref(A_j) = \texttt{MIN} 
	\end{array}
	\right.
\end{equation}

\begin{example}
	Avec la relation \texttt{Pokémon} (cf. tableau~\ref{tab:relation_exemple_3}), rechercher la séquence idéale de Pokémon dans un combat combine des conditions sur la \texttt{Rareté}, qui doit être la plus basse possible, la \texttt{Durée}, la plus courte possible, et le taux d'\texttt{Échec} le plus bas possible.
\end{example}

\subsubsection{IV. Scores avec le cosinus de Salton}\label{sssec:etape_4_calcul_des_scores_avec_le_cosinus_de_salton}

Cette étape vise à déterminer le score d'un point de Skyline avec le cosinus de Salton \footnote{Le cosinus de Salton, ou mesure de similarité cosinus, ou encore de cosinus de similitude, mesure, entre 0 à 1, la similarité entre vecteurs. Il permet de représenter une information par un vecteur et son importance par un angle dans un espace vectoriel. Il permet de calculer ainsi, classiquement, la pertinence d'une page Web pour une recherche donnée.}. Pour cela, le cosinus de l'angle entre le point idéal et le point du Skyline est calculé. Plus l'angle est faible (et donc le cosinus de l'angle est élevé), plus le point du Skyline est important.

Soit $S_{Score}$ l'ensemble des valeurs de scores des points du Skyline, le tuple $v_i  = (v_{i}[A_1], v_{i}[A_2], \dotsc, v_{i}[A_m]) \in S_P$ et le point idéal $I^+ = (I^+[A_1], I^+[A_2], \dotsc, I^+[A_m])$, alors nous avons~: 
\begin{equation}
	s_i = S_c(v_i, I^+) := cos(\theta) = \frac{v_i \cdot I^+}{||v_i|| \cdot ||I^+||}
\end{equation}
\begin{equation}\label{eq:cosinus_de_salton}
	s_i = \frac{\sum_{j=1}^m v_{i}[A_j] \cdot I^+[A_j]}{\sqrt{\sum_{j=1}^m v_{i}[A_j]^2} \cdot \sqrt{\sum_{j=1}^m I^+[A_j]^2}}, \forall s_i \in S_{Score}
\end{equation}

Une conséquence est que $s_i = 1$ si et seulement si le point du Skyline est considéré comme le plus intéressant, et $s_i = 0$ si et seulement s'il est considéré comme le moins intéressant.

Nous pouvons utiliser le principe de similarité de TOPSIS pour calculer le score de chaque point d'un Skyline de la manière suivante~: soit $I^-$ le point anti-idéal alors, $\forall v_i \in S_P$, si nous considérons $I^- = (I^-[A_1], I^-[A_2], \dotsc, I^-[A_m])$, nous avons~:
\begin{equation}
	I^-[A_j] = 
	\left\lbrace
	\begin{array}{l}
		\texttt{MAX}(v_i[A_j]) \equiv Pref(A_j) = \texttt{MIN} \\ 
		\texttt{MIN}(v_i[A_j]) \equiv Pref(A_j) = \texttt{MAX}
	\end{array}
	\right.
\end{equation}

\subsubsection{V. Classement des résultats}\label{sssec:etape_5_classement_des_resultats}

La dernière étape a pour objectif d'ordonner les points du Skyline en fonction des scores de manière décroissante.

\subsubsection{CoSky en SQL}

Il est à noter que de la méthode CoSky est complètement intégrable aux systèmes de gestion de bases de données (SGBD) relationnelles. Autrement dit, il est toujours possible d'utiliser une requête SQL pour la mettre en œuvre.

\begin{table}[htbp]
	\caption{La relation $\texttt{Pokémon}_{2}$}\label{tab:relation_exemple_3}
	\centering
	\begin{minipage}{\linewidth}
		\centering
		\small
		\begin{tabular}{c|ccc}
			\toprule
			\texttt{RowId} & \texttt{Rareté} & \texttt{Durée} & \texttt{Échec} \\
			\midrule
			$1$ & $5$ & $20$ & $1 / 70$ \\
			$2$ & $4$ & $60$ & $1 / 50$ \\
			$3$ & $5$ & $30$ & $1 / 60$ \\
			$4$ & $1$ & $80$ & $1 / 60$ \\
			$5$ & $5$ & $90$ & $1 / 40$ \\
			$6$ & $9$ & $30$ & $1 / 50$ \\
			$7$ & $7$ & $80$ & $1 / 40$ \\
			$8$ & $9$ & $90$ & $1 / 30$ \\
			\bottomrule
		\end{tabular}
	\end{minipage}
\end{table}

\begin{example}
	Il est possible d'appliquer les différentes étapes de calcul de la méthode CoSky à la relation \texttt{Pokémon} (cf. tableau~\ref{tab:relation_exemple_3}), avec préférences Skyline unifiées $(\texttt{MIN}, \texttt{MIN}, \texttt{MIN})$ en utilisant la requête SQL suivante\footnote{Si le Skyline n'a qu'un seul point, son classement est inutile, ou doit être fait en encadrant de $\texttt{COALESCE}(\texttt{NULLIF}(..., 0), 1)$ chaque dénominateur de la requête.}~:
	\lstset{style=SQLStyle}
	\begin{lstlisting}[ language=SQL,
						deletekeywords={IDENTITY},
						deletekeywords={[2]INT},
						morekeywords={CLUSTERED, SKYLINE, OF, SQRT, ROUND},
						framesep=8pt,
						xleftmargin=40pt,
						framexleftmargin=40pt,
						frame=tb,
						framerule=0pt ]
		WITH S AS (SELECT * FROM Pokémon
		    SKYLINE OF Rareté MIN, Durée MIN, Échec MIN
		), SN AS (SELECT RowId,
		       Rareté / TRare AS NRare,
		       Durée / TDurée AS NDurée,
		       Échec / TÉchec AS NÉchec 
		FROM S, (SELECT SUM(Rareté) AS TRare, 
		  SUM(Durée) AS TDurée, 
		  SUM(Échec) AS TÉchec FROM S) AS ST
		), SGini AS (SELECT 
		  1 - SUM(NRare * NRare) AS GRare,
		  1 - SUM(NDurée * NDurée) AS GDurée,
		  1 - SUM(NÉchec * NÉchec) AS GÉchec FROM SN
		), SW AS (SELECT 
		  GRare / (GRare + GDurée + GÉchec) AS WRare,
		  GDurée / (GRare + GDurée + GÉchec) AS WDurée,
		  GÉchec / (GRare + GDurée + GÉchec) AS WÉchec
		  FROM SGini
		), SP AS (SELECT RowId,  
		  WRare * NRare AS PRare,
		  WDurée * NDurée AS PDurée,
		  WÉchec * NÉchec AS PÉchec FROM SN, SW
		), Idéal AS (SELECT MIN(PRare) AS IRare,  
		  MIN(PDurée) AS IDurée, 
		  MAX(PÉchec) AS IÉchec FROM SP
		), SScore AS (SELECT RowId,
		  (IRare * PRare + IDurée * PDurée + 
		   IÉchec * PÉchec) / 
		  (SQRT(PRare * PRare + PDurée *
		   PDurée + PÉchec * PÉchec) *
		   SQRT(IRare * IRare + IDurée *
		   IDurée + IÉchec * IÉchec)) AS Score 
		  FROM Idéal, SP)
		SELECT P.RowId AS RowId, Rareté, Durée, Échec, 
		       ROUND(Score, 3) AS Score
		FROM S P INNER JOIN SScore rs 
		  ON P.RowId = rs.RowId
		ORDER BY Score DESC;
	\end{lstlisting}
	
	Les résultats obtenus sont donnés au tableau~\ref{tab:classement_de_skyline_avec_la_methode_CoSky}.
	
	\begin{table}[htb]
		\caption{Classement de Skyline avec CoSky}\label{tab:classement_de_skyline_avec_la_methode_CoSky}
		\centering
		\begin{minipage}{\linewidth}
			\centering
			\small
			\begin{tabular}{c|cccc}
				\toprule
				\texttt{RowId} & \texttt{Rareté} & \texttt{Durée} & \texttt{Échec} & \texttt{Score} \\
				\midrule
				$2$ & $4$ & $60$ & $1 / 50$ & $0.909$ \\
				$4$ & $1$ & $80$ & $1 / 60$ & $0.847$ \\
				$1$ & $5$ & $20$ & $1 / 70$ & $0.774$ \\
				\bottomrule
			\end{tabular}
		\end{minipage}
	\end{table}
	
	Nous voyons donc que, contrairement à dp-idp, CoSky permet de distinguer clairement les points de Skyline. Les points de Skyline de \texttt{RowId} $2$ et $4$ ont, avec CoSky, des scores différents de $0$ et différents l'un de l'autre, alors qu'ils étaient tous les deux de $0$ avec la méthode dp-idp. La méthode CoSky permet donc bien d'avoir un classement de Skyline complet. En outre, avec dp-idp, l'importance d'un point est inversement proportionnelle au nombre de points de Skyline qui le dominent, ce qui peut être contestable suivant les applications. CoSky permet de mesurer un écartement par rapport à un idéal, ce qui constitue une autre vision au moins aussi pertinente. Dans l'exemple, non seulement les points de Skyline de \texttt{RowId} $2$ et $4$ sont différenciés, mais en plus le classement n'est pas le même qu'avec la méthode dp-idp.
\end{example}

\subsubsection{Algorithme de CoSky}

Bien que l'intégrabilité de CoSky aux SGBD relationnelles soit une propriété avantageuse de la méthode, il peut arriver qu'une implémentation algorithmique soit préférable. En effet, dans les cas d'une cardinalité dimensionnelle importante ou volatile, ou encore d'une volonté d'intégration simplifiée au sein d'un ensemble algorithmique plus vaste, la généricité et l'adaptabilité offerte par une solution algorithmique peut être préférable.

Dans cette optique, nous proposons l'algorithme naïf\footnote{L'algorithme CoSky a été optimisé, mais il n'emploie pas de statistiques sur les données, ne tire pas partie du parallélisme ou de la gestion de caches, etc.}~\ref{algo:CoSky_1_2}.

L'algorithme CoSky ne calcule pas, à proprement parler, le Skyline. Il est donc nécessaire de faire appel à une solution externe pour cela. Assez classiquement, notre choix s'est porté sur l'algorithme branch-and-bound skyline (BBS) (\cite{papadiasProgressiveSkylineComputation2005}) pour son efficacité.

L'algorithme CoSky est, bien sûr, très proche de l'implémentation de CoSky en SQL. Les étapes de la méthodes sont respectées bien que, pour des raisons d'optimisation, et notamment le plus souvent de mutualisation de traitements ou d'itérations, elles peuvent parfois être scindées ou voir leur ordre être bouleversé.

Afin d'aider à rendre les étapes algorithmiques plus aisées à comprendre, nous les avons annoté d'exemples tirés directement de l'application de la méthode CoSky à la relation \texttt{Pokémon} (cf. tableau~\ref{tab:relation_exemple_3}), avec préférences Skyline unifiées $(\texttt{MIN}, \texttt{MIN}, \texttt{MIN})$.

\begin{algorithm}[htbp]
	\caption{Algorithme CoSky ($\mathcal{O}(|S| \cdot |\mathcal{D}|)$)\label{algo:CoSky_1_2}}
	\begin{algorithmic}
		\INPUT~ \\
		La relation $r$. \\
		\OUTPUT~ \\
		Le tableau des scores CoSky $score$. \\
		\LET $S$~: le Skyline de $r$
		\COMMENT{Calculé avec BBS}
		\LET $\mathcal{D} = \{d_1, \dotsc, d_n \}$~: l'ensemble des dimensions de $r$.
		\LET $sum\mathcal{D}_S$~: un tableau associatif de $|\mathcal{D}|$ entiers 0
		\LET $S_N$~: la future relation $S$ normalisée
		\LET $sum\mathcal{D}_{S_N}^2$~: un tableau associatif de $|\mathcal{D}|$ entiers 0
		\LET $gini$~: nouveau tableau associatif de $|\mathcal{D}|$ entiers
		\LET $sum_{gini} := 0$;
		\LET $S_{NP}$~: la future relation $S$ normalisée pondérée
		\LET $sum\mathcal{D}_{S_{NP}}^2$~: un tableau associatif de $|\mathcal{D}|$ entiers 0
		\LET $ideal$~: un tableau associatif de $|\mathcal{D}|$ entiers 1
		\LET $sum_{ideal}^2 := 0$;
		\LET $sqrt_{sum_{ideal}^2}$~: un entier positif
		\LET $score$~: un tableau associatif de $|S|$ entiers 0
		\FORALL{$t \in S$}
		\FORALL{$d_j \in \mathcal{D}$}
		\STATE $sum\mathcal{D}_S[d_j] := sum\mathcal{D}_S[d_j] + t[d_j]$;
		\COMMENT{\emph{i.e.} SUM(Durée)}
		\ENDFOR
		\ENDFOR
		\FORALL{$t \in S_N$}
		\LET $u$~: $u \in S, u[\texttt{RowId}] = t[\texttt{RowId}]$;
		\FORALL{$d_j \in \mathcal{D}$}
		\STATE $t[d_j] := u[d_j] / sum\mathcal{D}_S[d_j]$;
		\COMMENT{\emph{i.e.} Durée / TDurée}
		\STATE $sum\mathcal{D}_{S_N}^2[d_j] := sum\mathcal{D}_{S_N}^2[d_j] + t[d_j]^2$;
		\STATE
		\COMMENT{\emph{i.e.} SUM(NDurée * NDurée)}
		\ENDFOR
		\ENDFOR
		\FORALL{$d_j \in \mathcal{D}$}
		\STATE $gini[d_j] := 1 - sum\mathcal{D}_{S_N}^2[d_j]$;
		\STATE
		\COMMENT{\emph{i.e.} 1 - SUM(NDurée * NDurée)}
		\STATE $sum_{gini} := sum_{gini} + gini[d_j]$;
		\STATE 
		\COMMENT{\emph{i.e.} GRareté + GDurée + GÉchec}
		\ENDFOR
		\FORALL{$t \in S_{NP}$}
		\LET $i := t[\texttt{RowId}]$;
		\LET $u$~: $u \in S_N, u[\texttt{RowId}] = i$;
		\FORALL{$d_j \in \mathcal{D}$}
		\STATE $t[d_j] := gini[d_j] / sum_{gini} \times u[d_j]$;
		\STATE 
		\COMMENT{\emph{i.e.} GDurée / (GRareté + GDurée + GÉchec)\dots}
		\STATE 
		\COMMENT{\dots \  AS WRareté et WDurée * NDurée}
		\STATE $sum\mathcal{D}_{S_{NP}}^2[i] := sum\mathcal{D}_{S_{NP}}^2[i] + t[d_j]^2$;
		\STATE
		\COMMENT{\emph{i.e.} PRareté * PRareté + \dots \ + PÉchec * PÉchec}
		\IF{$t[d_j] < ideal[d_j]$}
		\STATE $ideal[d_j] := t[d_j]$;
		\COMMENT{\emph{i.e.} MIN(Durée)}
		\ENDIF
		\ENDFOR
		\ENDFOR
		\FORALL{$d_j \in \mathcal{D}$}
		\STATE $sum_{ideal}^2 := sum_{ideal}^2 + ideal[d_j]^2$;
		\STATE 
		\COMMENT{\emph{i.e.} IRareté * IRareté + \dots \ + IÉchec * IÉchec}
		\ENDFOR
		\STATE $sqrt_{sum_{ideal}^2} := \sqrt{sum_{ideal}^2}$;
		\FORALL{$t \in S_{NP}$}
		\LET $i := t[\texttt{RowId}]$;
		\LET $score_{numerator} := 0$;
		\FORALL{$d_j \in \mathcal{D}$}
		\STATE $score_{numerateur} := score_{numerateur} + ideal[d_j] \times t[d_j]$;
		\STATE 
		\COMMENT{\emph{i.e.} IRareté * PRareté + \dots \ + IÉchec * PÉchec}
		\ENDFOR
		\STATE $score[i] := score_{numerateur} / (\sqrt{sum\mathcal{D}_{S_{NP}}^2[i]} \times sqrt_{sum_{ideal}^2})$;
		\STATE 
		\COMMENT{cf. formule~\ref{eq:cosinus_de_salton}}
		\ENDFOR
		\RETURN $score$;
	\end{algorithmic}
\end{algorithm}

\subsection{Méthode top-$k$}\index{Top-$k$}

Pour cette dernière méthode, nous utilisons le principe de Skyline multiniveaux (\cite{preisingerLookingBestNot2015}) permettant de trouver les top-$k$ points de Skyline, non ordonnés entre eux. Une requête Skyline top-$k$ $Q_k$ sur une relation $r$ calcule les top-$k$ points, en fonction des préférences Skyline de $S$. Soit les points du niveau $0$ du Skyline multiniveaux $S_0(r)$, ou points du Skyline, tels que $S_0(r) = S$, et $Card(r)$ la cardinalité de $r$ telle que $Card(r) > k$, alors~:
\begin{itemize}
	\item si $Card(S_0(r)) > k$~: $Q_k$ ne renvoie que $k$ points de $S_0(r)$~;
	\item si $Card(S_0(r)) = k$~: $Q_k$ retourne le Skyline entier (\emph{i.e.} tous les points de $S_0(r)$)~;
	\item si $Card(S_0(r)) < k$~: il n'y a pas assez de points dans $S_0(r)$ pour permettre une réponse correcte avec $Q_k$. Une approche Skyline multiniveaux doit alors être appliquée. Cela signifie que, non seulement, des points de $S_1(r)$ de $(r \backslash S_0(r))$ sont retournés, mais aussi potentiellement certains points de $S_2(r)$ de $(r \backslash (S_0(r) \cup S_1(r))$, de $S_3(r)$\dots tant que le nombre cumulé de résultats retournés est inférieur à $k$.
\end{itemize}

\subsubsection{DeepSky}\index{DeepSky}

L'algorithme DeepSky (cf. algorithme~\ref{algo:deep_sky}) utilise ce principe multiniveaux allié à la méthode de classement CoSky afin de trouver les top-$k$ points de Skyline ordonnés. Il retourne les $k$ points de Skyline multiniveaux qui ont les $k$ plus haut scores calculés par la méthode CoSky.

\begin{algorithm}[htb]
	\caption{Algorithme DeepSky ($\mathcal{O}(k)$)\label{algo:deep_sky}}
	\begin{algorithmic}
		\INPUT~ \\
		La relation $r$. \\
		Le nombre $k$. \\
		\OUTPUT~ \\
		Les top-$k$ tuples/points avec les meilleurs scores $top_k$. \\
		\LET $top_k := \emptyset$;
		\LET $tot := 0$;
		\COMMENT{Nombre total de résultats calculés}
		\LET $r_l := r$;
		\COMMENT{Niveau courant}
		\WHILE{$tot < k \vee r_l = \emptyset$}
		\STATE $S := \texttt{CoSky}(r_l)$;
		\STATE $tot := tot + |S|$;
		\IF{$tot \le k$}
		\STATE $top_k := top_k \cup S$;
		\STATE $r_l := r_l \backslash S$;
		\ELSE
		\LET $S_trunc$~: les $k$ premiers points de $S$
		\STATE $top_k := top_k \cup S_trunc$;
		\RETURN $top_k$
		\ENDIF
		\ENDWHILE
		\RETURN $top_k$
	\end{algorithmic}
\end{algorithm}

\begin{example}
	Avec la relation \texttt{Pokémon} (cf. tableau~\ref{tab:relation_exemple_3}), et $k = 4$, l'algorithme DeepSky (cf. algorithme~\ref{algo:deep_sky}) renvoie les points de Skyline de \texttt{RowId} $1$, $4$ et $2$, les points de Skyline classés au niveau $0$, et le point de Skyline de \texttt{RowId} $3$, le seul point classé de niveau $1$.
\end{example}

\section{Discussion}\label{sec:discussion}

Le classement de Skyline par CoSky est avantageux, aussi bien par sa justesse que par son efficacité.

Le cosinus de Salton est une méthode puissante, flexible ainsi que simple à calculer et à interpréter. Il est également invariant aux transformations linéaires des vecteurs (comme la mise à l'échelle des valeurs) et est tout indiqué pour comparer des vecteurs dans des espaces de grande dimension, ce qui nous intéresse tout particulièrement en analyse de données. En revanche, il est sensible aux vecteurs nuls et ne prend pas en compte la magnitude des vecteurs, seulement leur direction. Ainsi, bien que rare, un point de Skyline composant un vecteur superposant le vecteur formé avec le point idéal, bien qu'il soit d'une amplitude plus forte, sera considéré, à tord, comme optimal. De même, plusieurs points peuvent avoir avoir la même mesure de similarité.

Nous avons considéré qu'un point idéal est optimal sur l'ensemble des dimensions. Nous pourrions comparer, plus finement, avec un point dominant tous les autres.

D'autres tentatives de tri de l'ensemble des points de Skyline ont été proposées dans la littérature avec l'objectif de maîtriser la taille du résultat. Parmi ces méthodes, celle utilisant la «~minimisation de regret~» (\cite{fabrisFlexibleSkylinesRegret2022}) pourrait aussi être comparée, ne serait-ce qu'expérimentalement, avec CoSky.

\section{Évaluations expérimentales}\label{sec:evaluations_experimentales}

Les expérimentations ont été réalisées sur une machine avec Intel(R) Xeon(R) W-11955M CPU @ 2.60GHz   2.61 GHz, avec 32Gb de mémoire RAM, fonctionnant sous Linux. Le code source a été écrit en Python 3.8 et interprété avec PyPy 3.9. PyPy est une implémentation alternative du langage de programmation Python, conçue pour être plus rapide et plus efficace en termes de consommation de mémoire par rapport à l'implémentation standard de Python CPython. En moyenne, PyPy 3.9 est 4.8 fois plus rapide que CPython 3.7. Les durées indiquées sont exprimées en secondes, mesurées en tant que temps de traitement du processeur et en supposant une valeur par défaut de 8~ms par défaut de page.

Spécifiquement concernant l'évaluation de l'algorithme SkyIR-UBS, les ensembles de données considérés ont été indexés avec agrégations dans un R*-arbre d'une taille de page de 4Ko. Un cache associé contenant 20~\% des blocs du R*-arbre correspondant a été utilisé lors de l'expérimentation.

Nous avons généré des ensembles réalistes et représentatifs de données synthétiques décorrélées composés de $10$ à $1$ milliard de tuples pour respectivement $3$, $6$ et $9$ dimensions reproduisant, chacune, des caractéristiques spécifiques (type, domaine de valeurs\dots) du cas d'utilisation.

\subsection{L'ensemble des solutions}\label{ssec:comparaison_de_l_ensemble_des_solutions}

Pour la comparaison avec l'ensemble des solutions, nous n'avons conservé que l'algorithme de dp-idp initial le plus performant, à savoir SkyIR-UBS.

\begin{figure}[htbp]
	\centering
	\resizebox{.45\linewidth}{!}{
		\begin{tikzpicture}[
			line join=bevel,
			biggraynode/.style={shape=circle, fill=gray, draw=black, line width=1pt},
			bigbrightmaroonnode/.style={shape=circle, fill=brightmaroon, draw=black, line width=1pt},
			bigcyannode/.style={shape=circle, fill=cyan, draw=black, line width=1pt},
			bigskybluenode/.style={shape=circle, fill=skyblue, draw=black, line width=1pt}
			]
			
			\draw[-stealth] (0pt, 0pt) -- (300pt, 0pt) node[anchor=north west] {Cardinalité};
			\draw[-stealth] (0pt, 0pt) -- (0pt, 340pt) node[anchor=south] {Temps de réponse en s};
			
			\foreach \y/\ytext in {0pt/$0$, 8pt/$600$, 40pt/$3000$, 80pt/$6000$, 120pt/$9000$, 160pt/$12000$, 200pt/$15000$, 240pt/$18000$, 280pt/$21000$, 320pt/$24000$} {
				\draw (2pt, \y) -- (-2pt, \y) node[left] {$\ytext\strut$};
			}
			\foreach \x/\xtext in {0pt/$0$, 56pt/$10000$, 112pt/$20000$, 168pt/$30000$, 224pt/$40000$, 280pt/$50000$} {
				\draw (\x, 2pt) -- (\x, -2pt) node[below] {$\xtext\strut$};
			}
			
			\draw[gray, line width=2pt](0pt, 0pt) -- (0pt, 0pt) -- (0pt, 0pt) -- (1pt, 0pt) -- (1pt, 0pt) -- (3pt, 0pt) -- (6pt, 0pt) -- (11pt, 1pt) -- (28pt, 7pt) -- (56pt, 25pt) -- (112pt, 138pt) -- (280pt, 280pt);
			\draw[brightmaroon, line width=2pt](0pt, 0pt) -- (0pt, 0pt) -- (0pt, 0pt) -- (1pt, 0pt) -- (1pt, 0pt) -- (3pt, 0pt) -- (6pt, 0pt) -- (11pt, 0pt) -- (28pt, 0pt) -- (56pt, 1pt) -- (112pt, 10pt) -- (280pt, 134pt);
			\draw[skyblue, line width=2pt](0pt, 0pt) -- (0pt, 0pt) -- (0pt, 0pt) -- (1pt, 0pt) -- (1pt, 0pt) -- (3pt, 0pt) -- (6pt, 0pt) -- (11pt, 0pt) -- (28pt, 0pt) -- (56pt, 0pt) -- (112pt, 0pt) -- (280pt, 3pt);
			\draw[cyan, line width=2pt](0pt, 0pt) -- (0pt, 0pt) -- (0pt, 0pt) -- (1pt, 0pt) -- (1pt, 0pt) -- (3pt, 0pt) -- (6pt, 0pt) -- (11pt, 0pt) -- (28pt, 0pt) -- (56pt, 0pt) -- (112pt, 0pt) -- (280pt, 0pt);
			
			\filldraw[color=black, fill=gray] (0pt, 0pt) circle (2pt);
			\filldraw[color=black, fill=gray] (0pt, 0pt) circle (2pt);
			\filldraw[color=black, fill=gray] (0pt, 0pt) circle (2pt);
			\filldraw[color=black, fill=gray] (1pt, 0pt) circle (2pt);
			\filldraw[color=black, fill=gray] (1pt, 0pt) circle (2pt);
			\filldraw[color=black, fill=gray] (3pt, 0pt) circle (2pt);
			\filldraw[color=black, fill=gray] (6pt, 0pt) circle (2pt);
			\filldraw[color=black, fill=gray] (11pt, 1pt) circle (2pt);
			\filldraw[color=black, fill=gray] (28pt, 7pt) circle (2pt);
			\filldraw[color=black, fill=gray] (56pt, 25pt) circle (2pt);
			\filldraw[color=black, fill=gray] (112pt, 138pt) circle (2pt);
			\filldraw[color=black, fill=gray] (280pt, 280pt) circle (2pt);
			\filldraw[color=black, fill=brightmaroon] (0pt, 0pt) circle (2pt);
			\filldraw[color=black, fill=brightmaroon] (0pt, 0pt) circle (2pt);
			\filldraw[color=black, fill=brightmaroon] (0pt, 0pt) circle (2pt);
			\filldraw[color=black, fill=brightmaroon] (1pt, 0pt) circle (2pt);
			\filldraw[color=black, fill=brightmaroon] (1pt, 0pt) circle (2pt);
			\filldraw[color=black, fill=brightmaroon] (3pt, 0pt) circle (2pt);
			\filldraw[color=black, fill=brightmaroon] (6pt, 0pt) circle (2pt);
			\filldraw[color=black, fill=brightmaroon] (11pt, 0pt) circle (2pt);        
			\filldraw[color=black, fill=brightmaroon] (28pt, 0pt) circle (2pt);        
			\filldraw[color=black, fill=brightmaroon] (56pt, 1pt) circle (2pt);        
			\filldraw[color=black, fill=brightmaroon] (112pt, 10pt) circle (2pt);       
			\filldraw[color=black, fill=brightmaroon] (280pt, 134pt) circle (2pt);
			\filldraw[color=black, fill=skyblue] (0pt, 0pt) circle (2pt);
			\filldraw[color=black, fill=skyblue] (0pt, 0pt) circle (2pt);
			\filldraw[color=black, fill=skyblue] (0pt, 0pt) circle (2pt);
			\filldraw[color=black, fill=skyblue] (1pt, 0pt) circle (2pt);
			\filldraw[color=black, fill=skyblue] (1pt, 0pt) circle (2pt);
			\filldraw[color=black, fill=skyblue] (3pt, 0pt) circle (2pt);
			\filldraw[color=black, fill=skyblue] (6pt, 0pt) circle (2pt);
			\filldraw[color=black, fill=skyblue] (11pt, 0pt) circle (2pt);
			\filldraw[color=black, fill=skyblue] (28pt, 0pt) circle (2pt);
			\filldraw[color=black, fill=skyblue] (56pt, 0pt) circle (2pt);
			\filldraw[color=black, fill=skyblue] (112pt, 0pt) circle (2pt);
			\filldraw[color=black, fill=skyblue] (280pt, 3pt) circle (2pt);   
			\filldraw[color=black, fill=cyan] (0pt, 0pt) circle (2pt);
			\filldraw[color=black, fill=cyan] (0pt, 0pt) circle (2pt);
			\filldraw[color=black, fill=cyan] (0pt, 0pt) circle (2pt);
			\filldraw[color=black, fill=cyan] (1pt, 0pt) circle (2pt);
			\filldraw[color=black, fill=cyan] (1pt, 0pt) circle (2pt);
			\filldraw[color=black, fill=cyan] (3pt, 0pt) circle (2pt);
			\filldraw[color=black, fill=cyan] (6pt, 0pt) circle (2pt);
			\filldraw[color=black, fill=cyan] (11pt, 0pt) circle (2pt);
			\filldraw[color=black, fill=cyan] (28pt, 0pt) circle (2pt);
			\filldraw[color=black, fill=cyan] (56pt, 0pt) circle (2pt);
			\filldraw[color=black, fill=cyan] (112pt, 0pt) circle (2pt);
			\filldraw[color=black, fill=cyan] (280pt, 0pt) circle (2pt);
			
			\matrix [below left] at (current bounding box.north east) {
				\node [biggraynode, label=right:SkyIR-UBS] {}; \\
				\node [bigbrightmaroonnode, label=right:dp-idp avec hiérarchie de dominance] {}; \\
				\node [bigskybluenode, label=right:CoSky «~Algorithme~»] {}; \\
				\node [bigcyannode, label=right:CoSky «~SQL~»] {}; \\
			};
		\end{tikzpicture}
	}
	\caption{Temps de réponse des différentes solutions}\label{fig:temps_de_reponse_des_differentes_solutions}
\end{figure}
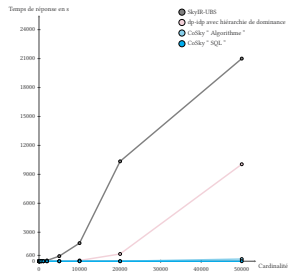

Même à cette condition, la figure~\ref{fig:temps_de_reponse_des_differentes_solutions}, montrant le temps de réponses des différentes solutions, lorsque la cardinalité de l'ensemble des données varie jusqu'à $50000$ tuples, pour $3$ dimensions, évalue SkyIR-UBS comme la moins efficace des solutions. Notre proposition de dp-idp avec hiérarchie de dominance explose nettement moins vite. Pourtant, bien que meilleure que celle de SkyIR-UBS, son efficacité est négligeable face aux implémentations de CoSky. Même dans le pire cas ($50000$ tuples et $3$ dimensions), les implémentations SQL et algorithmique de CoSky ont respectivement un temps de réponse de $0.268$ seconde et de $4$ minutes et $23$ secondes là où notre version de dp-idp et SkyIR-UBS en ont un respectif de plus de $3$ heures et de plus de $6$ heures~!

Pour la suite des évaluations, nous délaissons les algorithmes SkyIR-UBS et dp-idp avec hiérarchie de dominance, peu efficaces par rapport aux implémentations de CoSky, afin de pouvoir les étudier avec de plus fortes cardinalités et de plus grandes dimensionnalités.

\subsection{Implémentations de CoSky}\label{ssec:comparaison_des_implementations_de_CoSky}

\begin{figure}[htbp]
	\centering
	\resizebox{.45\linewidth}{!}{
		\begin{tikzpicture}[
			line join=bevel,
			bigcyannode/.style={shape=circle, fill=cyan, draw=black, line width=1pt},
			bigskybluenode/.style={shape=circle, fill=skyblue, draw=black, line width=1pt}
			]
			
			\draw[-stealth] (0pt, 0pt) -- (300pt, 0pt) node[anchor=north west] {Cardinalité};
			\draw[-stealth] (0pt, 0pt) -- (0pt, 310pt) node[anchor=south] {Temps de réponse en s};
			
			\foreach \y/\ytext in {0pt/$0$, 9pt/$60$, 55pt/$500$, 110pt/$1000$, 165pt/$1500$, 220pt/$2000$, 275pt/$2500$} {
				\draw (2pt, \y) -- (-2pt, \y) node[left] {$\ytext\strut$};
			}
			\foreach \x/\xtext in {0pt/$0$, 56pt/$40000$, 112pt/$80000$, 168pt/$120000$, 224pt/$160000$, 280pt/$200000$} {
				\draw (\x, 2pt) -- (\x, -2pt) node[below] {$\xtext\strut$};
			}
			
			\draw[cyan, line width=2pt](0pt, 0pt) -- (0pt, 0pt) -- (0pt, 0pt) -- (0pt, 0pt) -- (0pt, 0pt) -- (1pt, 0pt) -- (1pt, 0pt) -- (3pt, 0pt) -- (7pt, 0pt) -- (14pt, 0pt) -- (28pt, 0pt) -- (70pt, 0pt) -- (140pt, 0pt) -- (280pt, 0pt);
			\draw[skyblue, line width=2pt](0pt, 0pt) -- (0pt, 0pt) -- (0pt, 0pt) -- (0pt, 0pt) -- (0pt, 0pt) -- (1pt, 0pt) -- (1pt, 0pt) -- (3pt, 0pt) -- (7pt, 0pt) -- (14pt, 0pt) -- (28pt, 2pt) -- (70pt, 13pt) -- (140pt, 80pt) -- (280pt, 280pt);
			
			\filldraw[color=black, fill=cyan] (0pt, 0pt) circle (2pt);
			\filldraw[color=black, fill=cyan] (0pt, 0pt) circle (2pt);
			\filldraw[color=black, fill=cyan] (0pt, 0pt) circle (2pt);
			\filldraw[color=black, fill=cyan] (0pt, 0pt) circle (2pt);
			\filldraw[color=black, fill=cyan] (0pt, 0pt) circle (2pt);
			\filldraw[color=black, fill=cyan] (1pt, 0pt) circle (2pt);
			\filldraw[color=black, fill=cyan] (1pt, 0pt) circle (2pt);
			\filldraw[color=black, fill=cyan] (3pt, 0pt) circle (2pt);
			\filldraw[color=black, fill=cyan] (7pt, 0pt) circle (2pt);
			\filldraw[color=black, fill=cyan] (14pt, 0pt) circle (2pt);
			\filldraw[color=black, fill=cyan] (28pt, 0pt) circle (2pt);
			\filldraw[color=black, fill=cyan] (70pt, 0pt) circle (2pt);
			\filldraw[color=black, fill=cyan] (140pt, 0pt) circle (2pt);
			\filldraw[color=black, fill=cyan] (280pt, 0pt) circle (2pt);
			\filldraw[color=black, fill=skyblue] (0pt, 0pt) circle (2pt);
			\filldraw[color=black, fill=skyblue] (0pt, 0pt) circle (2pt);
			\filldraw[color=black, fill=skyblue] (0pt, 0pt) circle (2pt);
			\filldraw[color=black, fill=skyblue] (0pt, 0pt) circle (2pt);
			\filldraw[color=black, fill=skyblue] (0pt, 0pt) circle (2pt);
			\filldraw[color=black, fill=skyblue] (1pt, 0pt) circle (2pt);
			\filldraw[color=black, fill=skyblue] (1pt, 0pt) circle (2pt);
			\filldraw[color=black, fill=skyblue] (3pt, 0pt) circle (2pt);
			\filldraw[color=black, fill=skyblue] (7pt, 0pt) circle (2pt);
			\filldraw[color=black, fill=skyblue] (14pt, 0pt) circle (2pt);
			\filldraw[color=black, fill=skyblue] (28pt, 2pt) circle (2pt);
			\filldraw[color=black, fill=skyblue] (70pt, 13pt) circle (2pt);
			\filldraw[color=black, fill=skyblue] (140pt, 80pt) circle (2pt);
			\filldraw[color=black, fill=skyblue] (280pt, 280pt) circle (2pt);
			
			\matrix [below left] at (current bounding box.north east) {
				\node [bigskybluenode, label=right:CoSky «~Algorithme~» avec 3 attributs] {}; \\
				\node [bigcyannode, label=right:CoSky «~SQL~» avec 3 attributs] {}; \\
			};
		\end{tikzpicture}
		}
		\resizebox{.45\linewidth}{!}{
		\begin{tikzpicture}[
			line join=bevel,
			bigcyannode/.style={shape=circle, fill=cyan, draw=black, line width=1pt},
			bigskybluenode/.style={shape=circle, fill=skyblue, draw=black, line width=1pt}
			]
			
			\draw[-stealth] (0pt, 0pt) -- (300pt, 0pt) node[anchor=north west] {Cardinalité};
			\draw[-stealth] (0pt, 0pt) -- (0pt, 310pt) node[anchor=south] {Temps de réponse en s};
			
			\foreach \y/\ytext in {0pt/$0$, 8pt/$90$, 43pt/$500$, 85pt/$1000$, 170pt/$2000$, 255pt/$3000$} {
				\draw (2pt, \y) -- (-2pt, \y) node[left] {$\ytext\strut$};
			}
			\foreach \x/\xtext in {0pt/$0$, 56pt/$40000$, 112pt/$80000$, 168pt/$120000$, 224pt/$160000$, 280pt/$200000$} {
				\draw (\x, 2pt) -- (\x, -2pt) node[below] {$\xtext\strut$};
			}
			
			\draw[skyblue, line width=2pt](0pt, 0pt) -- (0pt, 0pt) -- (0pt, 0pt) -- (0pt, 0pt) -- (0pt, 0pt) -- (1pt, 0pt) -- (1pt, 0pt) -- (3pt, 0pt) -- (7pt, 0pt) -- (14pt, 0pt) -- (28pt, 2pt) -- (70pt, 11pt) -- (140pt, 59pt) -- (280pt, 280pt);
			\draw[cyan, line width=2pt](0pt, 0pt) -- (0pt, 0pt) -- (0pt, 0pt) -- (0pt, 0pt) -- (0pt, 0pt) -- (1pt, 0pt) -- (1pt, 0pt) -- (3pt, 0pt) -- (7pt, 0pt) -- (14pt, 0pt) -- (28pt, 0pt) -- (70pt, 1pt) -- (140pt, 2pt) -- (280pt, 5pt);
			
			\filldraw[color=black, fill=skyblue] (0pt, 0pt) circle (2pt);
			\filldraw[color=black, fill=skyblue] (0pt, 0pt) circle (2pt);
			\filldraw[color=black, fill=skyblue] (0pt, 0pt) circle (2pt);
			\filldraw[color=black, fill=skyblue] (0pt, 0pt) circle (2pt);
			\filldraw[color=black, fill=skyblue] (0pt, 0pt) circle (2pt);
			\filldraw[color=black, fill=skyblue] (1pt, 0pt) circle (2pt);
			\filldraw[color=black, fill=skyblue] (1pt, 0pt) circle (2pt);
			\filldraw[color=black, fill=skyblue] (3pt, 0pt) circle (2pt);
			\filldraw[color=black, fill=skyblue] (7pt, 0pt) circle (2pt);
			\filldraw[color=black, fill=skyblue] (14pt, 0pt) circle (2pt);
			\filldraw[color=black, fill=skyblue] (28pt, 2pt) circle (2pt);
			\filldraw[color=black, fill=skyblue] (70pt, 11pt) circle (2pt);
			\filldraw[color=black, fill=skyblue] (140pt, 59pt) circle (2pt);
			\filldraw[color=black, fill=skyblue] (280pt, 280pt) circle (2pt);
			\filldraw[color=black, fill=cyan] (0pt, 0pt) circle (2pt);
			\filldraw[color=black, fill=cyan] (0pt, 0pt) circle (2pt);
			\filldraw[color=black, fill=cyan] (0pt, 0pt) circle (2pt);
			\filldraw[color=black, fill=cyan] (0pt, 0pt) circle (2pt);
			\filldraw[color=black, fill=cyan] (0pt, 0pt) circle (2pt);
			\filldraw[color=black, fill=cyan] (1pt, 0pt) circle (2pt);
			\filldraw[color=black, fill=cyan] (1pt, 0pt) circle (2pt);
			\filldraw[color=black, fill=cyan] (3pt, 0pt) circle (2pt);
			\filldraw[color=black, fill=cyan] (7pt, 0pt) circle (2pt);
			\filldraw[color=black, fill=cyan] (14pt, 0pt) circle (2pt);
			\filldraw[color=black, fill=cyan] (28pt, 0pt) circle (2pt);
			\filldraw[color=black, fill=cyan] (70pt, 1pt) circle (2pt);
			\filldraw[color=black, fill=cyan] (140pt, 2pt) circle (2pt);
			\filldraw[color=black, fill=cyan] (280pt, 5pt) circle (2pt);
			
			\matrix [below left] at (current bounding box.north east) {
				\node [bigskybluenode, label=right:CoSky «~Algorithme~» avec 6 attributs] {}; \\
				\node [bigcyannode, label=right:CoSky «~SQL~» avec 6 attributs] {}; \\
			};
		\end{tikzpicture}
	}
	\resizebox{.45\linewidth}{!}{
		\begin{tikzpicture}[
			line join=bevel,
			bigcyannode/.style={shape=circle, fill=cyan, draw=black, line width=1pt},
			bigskybluenode/.style={shape=circle, fill=skyblue, draw=black, line width=1pt}
			]
			
			\draw[-stealth] (0pt, 0pt) -- (300pt, 0pt) node[anchor=north west] {Cardinalité};
			\draw[-stealth] (0pt, 0pt) -- (0pt, 310pt) node[anchor=south] {Temps de réponse en s};
			
			\foreach \y/\ytext in {0pt/$0$, 28pt/$500$, 55pt/$1000$, 110pt/$2000$, 165pt/$3000$, 220pt/$4000$, 275pt/$5000$} {
				\draw (2pt, \y) -- (-2pt, \y) node[left] {$\ytext\strut$};
			}
			\foreach \x/\xtext in {0pt/$0$, 56pt/$40000$, 112pt/$80000$, 168pt/$120000$, 224pt/$160000$, 280pt/$200000$} {
				\draw (\x, 2pt) -- (\x, -2pt) node[below] {$\xtext\strut$};
			}
			
			\draw[skyblue, line width=2pt](0pt, 0pt) -- (0pt, 0pt) -- (0pt, 0pt) -- (0pt, 0pt) -- (0pt, 0pt) -- (1pt, 0pt) -- (1pt, 0pt) -- (3pt, 0pt) -- (7pt, 0pt) -- (14pt, 0pt) -- (28pt, 2pt) -- (70pt, 9pt) -- (140pt, 81pt) -- (280pt, 280pt);
			\draw[cyan, line width=2pt](0pt, 0pt) -- (0pt, 0pt) -- (0pt, 0pt) -- (0pt, 0pt) -- (0pt, 0pt) -- (1pt, 0pt) -- (1pt, 0pt) -- (3pt, 0pt) -- (7pt, 0pt) -- (14pt, 0pt) -- (28pt, 0pt) -- (70pt, 2pt) -- (140pt, 7pt) -- (280pt, 23pt);
			
			\filldraw[color=black, fill=skyblue] (0pt, 0pt) circle (2pt);
			\filldraw[color=black, fill=skyblue] (0pt, 0pt) circle (2pt);
			\filldraw[color=black, fill=skyblue] (0pt, 0pt) circle (2pt);
			\filldraw[color=black, fill=skyblue] (0pt, 0pt) circle (2pt);
			\filldraw[color=black, fill=skyblue] (0pt, 0pt) circle (2pt);
			\filldraw[color=black, fill=skyblue] (1pt, 0pt) circle (2pt);
			\filldraw[color=black, fill=skyblue] (1pt, 0pt) circle (2pt);
			\filldraw[color=black, fill=skyblue] (3pt, 0pt) circle (2pt);
			\filldraw[color=black, fill=skyblue] (7pt, 0pt) circle (2pt);
			\filldraw[color=black, fill=skyblue] (14pt, 0pt) circle (2pt);
			\filldraw[color=black, fill=skyblue] (28pt, 2pt) circle (2pt);
			\filldraw[color=black, fill=skyblue] (70pt, 9pt) circle (2pt);
			\filldraw[color=black, fill=skyblue] (140pt, 81pt) circle (2pt);
			\filldraw[color=black, fill=skyblue] (280pt, 280pt) circle (2pt);
			\filldraw[color=black, fill=cyan] (0pt, 0pt) circle (2pt);
			\filldraw[color=black, fill=cyan] (0pt, 0pt) circle (2pt);
			\filldraw[color=black, fill=cyan] (0pt, 0pt) circle (2pt);
			\filldraw[color=black, fill=cyan] (0pt, 0pt) circle (2pt);
			\filldraw[color=black, fill=cyan] (0pt, 0pt) circle (2pt);
			\filldraw[color=black, fill=cyan] (1pt, 0pt) circle (2pt);
			\filldraw[color=black, fill=cyan] (1pt, 0pt) circle (2pt);
			\filldraw[color=black, fill=cyan] (3pt, 0pt) circle (2pt);
			\filldraw[color=black, fill=cyan] (7pt, 0pt) circle (2pt);
			\filldraw[color=black, fill=cyan] (14pt, 0pt) circle (2pt);
			\filldraw[color=black, fill=cyan] (28pt, 0pt) circle (2pt);
			\filldraw[color=black, fill=cyan] (70pt, 2pt) circle (2pt);
			\filldraw[color=black, fill=cyan] (140pt, 7pt) circle (2pt);
			\filldraw[color=black, fill=cyan] (280pt, 23pt) circle (2pt);
			
			\matrix [below left] at (current bounding box.north east) {
				\node [bigskybluenode, label=right:CoSky «~Algorithme~» avec 9 attributs] {}; \\
				\node [bigcyannode, label=right:CoSky «~SQL~» avec 9 attributs] {}; \\
			};
		\end{tikzpicture}
	}
	\caption{Temps de réponse de CoSky SQL (3, 6 et 9 attributs)}\label{fig:temps_de_reponse_des_requetes_CoSky}
\end{figure}
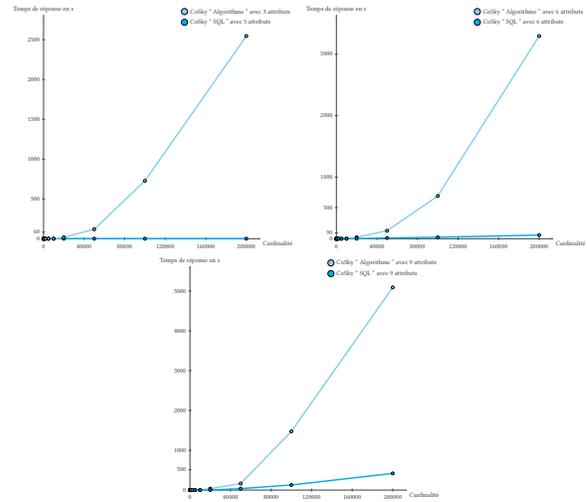

Pour la comparaison des implémentations de CoSky, nous considérons la  figure~\ref{fig:temps_de_reponse_des_requetes_CoSky}. Les évaluations sont effectuées pour des cardinalités allant jusqu'à $200000$ tuples avec, respectivement, $3$, $6$ et $9$ dimensions. Dans le pire des cas, les temps de réponse sont respectivement d'environ $40$ minutes, $53$ minutes et $1$ heure $43$ minutes pour l'implémentation algorithmique de CoSky alors qu'elles sont respectivement d'environ $1$ seconde, $54$ secondes et $2$ minutes et $8$ secondes pour l'implémentation SQL. Nous constatons que, bien que les deux solutions soient particulièrement efficaces, et notamment par rapport aux solutions existantes, la version intégrée au SGBD relationnelles l'est considérablement.

\subsection{CoSky en SQL}\label{ssec:comparaison_de_l_implementation_sql_de_CoSky}

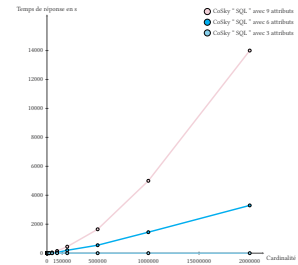
\begin{figure}[htbp]
	\centering
	\resizebox{.45\linewidth}{!}{
		\begin{tikzpicture}[
			line join=bevel,
			bigbrightmaroonnode/.style={shape=circle, fill=brightmaroon, draw=black, line width=1pt},
			bigcyannode/.style={shape=circle, fill=cyan, draw=black, line width=1pt},
			bigskybluenode/.style={shape=circle, fill=skyblue, draw=black, line width=1pt}
			]
			
			\draw[-stealth] (0pt, 0pt) -- (300pt, 0pt) node[anchor=north west] {Cardinalité};
			\draw[-stealth] (0pt, 0pt) -- (0pt, 330pt) node[anchor=south] {Temps de réponse en s};
			
			\foreach \y/\ytext in {0pt/$0$, 40pt/$2000$, 80pt/$4000$, 120pt/$6000$, 160pt/$8000$, 200pt/$10000$, 240pt/$12000$, 280pt/$14000$} {
				\draw (2pt, \y) -- (-2pt, \y) node[left] {$\ytext\strut$};
			}
			\foreach \x/\xtext in {0pt/$0$, 21pt/$150000$, 70pt/$500000$, 140pt/$1000000$, 210pt/$15000000$, 280pt/$2000000$} {
				\draw (\x, 2pt) -- (\x, -2pt) node[below] {$\xtext\strut$};
			}
			
			\draw[brightmaroon, line width=2pt](0pt, 0pt) -- (0pt, 0pt) -- (0pt, 0pt) -- (0pt, 0pt) -- (0pt, 0pt) -- (0pt, 0pt) -- (0pt, 0pt) -- (0pt, 0pt) -- (1pt, 0pt) -- (1pt, 0pt) -- (3pt, 0pt) -- (7pt, 1pt) -- (14pt, 3pt) -- (28pt, 9pt) -- (70pt, 33pt) -- (140pt, 100pt) -- (280pt, 280pt);     
			\draw[cyan, line width=2pt](0pt, 0pt) -- (0pt, 0pt) -- (0pt, 0pt) -- (0pt, 0pt) -- (0pt, 0pt) -- (0pt, 0pt) -- (0pt, 0pt) -- (0pt, 0pt) -- (1pt, 0pt) -- (1pt, 0pt) -- (3pt, 0pt) -- (7pt, 0pt) -- (14pt, 1pt) -- (28pt, 4pt) -- (70pt, 11pt) -- (140pt, 29pt) -- (280pt, 66pt);
			\draw[skyblue, line width=2pt](0pt, 0pt) -- (0pt, 0pt) -- (0pt, 0pt) -- (0pt, 0pt) -- (0pt, 
			0pt) -- (0pt, 0pt) -- (0pt, 0pt) -- (0pt, 0pt) -- (1pt, 0pt) -- (1pt, 0pt) -- (3pt, 0pt) -- (7pt, 0pt) -- (14pt, 0pt) -- (28pt, 0pt) -- (70pt, 0pt) -- (140pt, 0pt) -- (280pt, 0pt);
			
			\filldraw[color=black, fill=brightmaroon] (0pt, 0pt) circle (2pt);
			\filldraw[color=black, fill=brightmaroon] (0pt, 0pt) circle (2pt);
			\filldraw[color=black, fill=brightmaroon] (0pt, 0pt) circle (2pt);
			\filldraw[color=black, fill=brightmaroon] (0pt, 0pt) circle (2pt);
			\filldraw[color=black, fill=brightmaroon] (0pt, 0pt) circle (2pt);
			\filldraw[color=black, fill=brightmaroon] (0pt, 0pt) circle (2pt);
			\filldraw[color=black, fill=brightmaroon] (0pt, 0pt) circle (2pt);
			\filldraw[color=black, fill=brightmaroon] (0pt, 0pt) circle (2pt);
			\filldraw[color=black, fill=brightmaroon] (1pt, 0pt) circle (2pt);
			\filldraw[color=black, fill=brightmaroon] (1pt, 0pt) circle (2pt);
			\filldraw[color=black, fill=brightmaroon] (3pt, 0pt) circle (2pt);
			\filldraw[color=black, fill=brightmaroon] (7pt, 1pt) circle (2pt);
			\filldraw[color=black, fill=brightmaroon] (14pt, 3pt) circle (2pt);
			\filldraw[color=black, fill=brightmaroon] (28pt, 9pt) circle (2pt);
			\filldraw[color=black, fill=brightmaroon] (70pt, 33pt) circle (2pt);
			\filldraw[color=black, fill=brightmaroon] (140pt, 100pt) circle (2pt);
			\filldraw[color=black, fill=brightmaroon] (280pt, 280pt) circle (2pt);
			\filldraw[color=black, fill=cyan] (0pt, 0pt) circle (2pt);
			\filldraw[color=black, fill=cyan] (0pt, 0pt) circle (2pt);
			\filldraw[color=black, fill=cyan] (0pt, 0pt) circle (2pt);
			\filldraw[color=black, fill=cyan] (0pt, 0pt) circle (2pt);
			\filldraw[color=black, fill=cyan] (0pt, 0pt) circle (2pt);
			\filldraw[color=black, fill=cyan] (0pt, 0pt) circle (2pt);
			\filldraw[color=black, fill=cyan] (0pt, 0pt) circle (2pt);
			\filldraw[color=black, fill=cyan] (0pt, 0pt) circle (2pt);
			\filldraw[color=black, fill=cyan] (1pt, 0pt) circle (2pt);
			\filldraw[color=black, fill=cyan] (1pt, 0pt) circle (2pt);
			\filldraw[color=black, fill=cyan] (3pt, 0pt) circle (2pt);
			\filldraw[color=black, fill=cyan] (7pt, 0pt) circle (2pt);
			\filldraw[color=black, fill=cyan] (14pt, 1pt) circle (2pt);
			\filldraw[color=black, fill=cyan] (28pt, 4pt) circle (2pt);
			\filldraw[color=black, fill=cyan] (70pt, 11pt) circle (2pt);
			\filldraw[color=black, fill=cyan] (140pt, 29pt) circle (2pt);
			\filldraw[color=black, fill=cyan] (280pt, 66pt) circle (2pt);
			\filldraw[color=black, fill=skyblue] (0pt, 0pt) circle (2pt);
			\filldraw[color=black, fill=skyblue] (0pt, 0pt) circle (2pt);
			\filldraw[color=black, fill=skyblue] (0pt, 0pt) circle (2pt);
			\filldraw[color=black, fill=skyblue] (0pt, 0pt) circle (2pt);
			\filldraw[color=black, fill=skyblue] (0pt, 0pt) circle (2pt);
			\filldraw[color=black, fill=skyblue] (0pt, 0pt) circle (2pt);
			\filldraw[color=black, fill=skyblue] (0pt, 0pt) circle (2pt);
			\filldraw[color=black, fill=skyblue] (0pt, 0pt) circle (2pt);
			\filldraw[color=black, fill=skyblue] (1pt, 0pt) circle (2pt);
			\filldraw[color=black, fill=skyblue] (1pt, 0pt) circle (2pt);
			\filldraw[color=black, fill=skyblue] (3pt, 0pt) circle (2pt);
			\filldraw[color=black, fill=skyblue] (7pt, 0pt) circle (2pt);
			\filldraw[color=black, fill=skyblue] (14pt, 0pt) circle (2pt);
			\filldraw[color=black, fill=skyblue] (28pt, 0pt) circle (2pt);
			\filldraw[color=black, fill=skyblue] (70pt, 0pt) circle (2pt);
			\filldraw[color=black, fill=skyblue] (140pt, 0pt) circle (2pt);
			\filldraw[color=black, fill=skyblue] (280pt, 0pt) circle (2pt);
			
			\matrix [below left] at (current bounding box.north east) {
				\node [bigbrightmaroonnode, label=right:CoSky «~SQL~» avec 9 attributs] {}; \\
				\node [bigcyannode, label=right:CoSky «~SQL~» avec 6 attributs] {}; \\
				\node [bigskybluenode, label=right:CoSky «~SQL~» avec 3 attributs] {}; \\
			};
		\end{tikzpicture}
	}
	\caption{Temps de réponse de CoSky en SQL}\label{fig:temps_de_reponse_de_CoSky_en_sql}
\end{figure}

Pour la comparaison de l'implémentation SQL de CoSky, nous considérons la figure~\ref{fig:temps_de_reponse_de_CoSky_en_sql}. Les évaluations sont effectuées pour des cardinalités allant jusqu'à $2$ millions de tuples avec $3$, $6$ et $9$ dimensions. Dans le pire des cas, les temps de réponse sont d'environ $3$ heures pour $9$ dimensions, $16$ minutes pour $6$ dimensions et $10$ secondes pour $3$ colonnes. Comme attendu pour toutes sortes de calculs dans un contexte Skyline, notre implémentation SQL de CoSky est sensible à l'augmentation du nombre de dimensions. Ce problème est connu pour être difficile pour de hautes dimensionnalités, même avec une modélisation en RAM (\cite{chanHighDimensionalSkylines2006}). 

\subsection{CoSky avec 3 dimensions}\label{ssec:evaluation_de_l_implementation_sql_de_CoSky_avec_3_dimensions}

\begin{figure}[htbp]
	\centering
	\resizebox{.45\linewidth}{!}{
		\begin{tikzpicture}[
			line join=bevel,
			bigskybluenode/.style={shape=circle, fill=skyblue, draw=black, line width=1pt}
			]
			
			\draw[-stealth] (0pt, 0pt) -- (300pt, 0pt) node[anchor=north west] {Cardinalité};
			\draw[-stealth] (0pt, 0pt) -- (0pt, 300pt) node[anchor=south] {Temps de réponse en s};
			
			\foreach \y/\ytext in {0pt/$0$, 16pt/$600$, 40pt/$1500$, 80pt/$3000$, 120pt/$4500$, 160pt/$6000$, 200pt/$7500$, 240pt/$9000$, 280pt/$10500$} {
				\draw (2pt, \y) -- (-2pt, \y) node[left] {$\ytext\strut$};
			}
			\foreach \x/\xtext in {0pt/0, 56pt/200 \times 10^6, 112pt/400 \times 10^6, 168pt/600 \times 10^6, 224pt/800 \times 10^6, 280pt/10^9} {
				\draw (\x, 2pt) -- (\x, -2pt) node[below] {$\xtext\strut$};
			}
			
			\draw[skyblue, line width=2pt](0pt, 0pt) -- (0pt, 0pt) -- (0pt, 0pt) -- (0pt, 0pt) -- (0pt, 0pt) -- (0pt, 0pt) -- (0pt, 0pt) -- (0pt, 0pt) -- (0pt, 0pt) -- (0pt, 0pt) -- (0pt, 0pt) -- (0pt, 0pt) -- (0pt, 0pt) -- (0pt, 0pt) -- (0pt, 0pt) -- (0pt, 0pt) -- (1pt, 0pt) -- (1pt, 1pt) -- (3pt, 3pt) -- (6pt, 5pt) -- (14pt, 12pt) -- (28pt, 24pt) -- (56pt, 63pt) -- (140pt, 129pt) -- (280pt, 280pt);
			
			\filldraw[color=black, fill=skyblue] (0pt, 0pt) circle (2pt);
			\filldraw[color=black, fill=skyblue] (0pt, 0pt) circle (2pt);
			\filldraw[color=black, fill=skyblue] (0pt, 0pt) circle (2pt);
			\filldraw[color=black, fill=skyblue] (0pt, 0pt) circle (2pt);
			\filldraw[color=black, fill=skyblue] (0pt, 0pt) circle (2pt);
			\filldraw[color=black, fill=skyblue] (0pt, 0pt) circle (2pt);
			\filldraw[color=black, fill=skyblue] (0pt, 0pt) circle (2pt);
			\filldraw[color=black, fill=skyblue] (0pt, 0pt) circle (2pt);
			\filldraw[color=black, fill=skyblue] (0pt, 0pt) circle (2pt);
			\filldraw[color=black, fill=skyblue] (0pt, 0pt) circle (2pt);
			\filldraw[color=black, fill=skyblue] (0pt, 0pt) circle (2pt);
			\filldraw[color=black, fill=skyblue] (0pt, 0pt) circle (2pt);
			\filldraw[color=black, fill=skyblue] (0pt, 0pt) circle (2pt);
			\filldraw[color=black, fill=skyblue] (0pt, 0pt) circle (2pt);
			\filldraw[color=black, fill=skyblue] (0pt, 0pt) circle (2pt);
			\filldraw[color=black, fill=skyblue] (0pt, 0pt) circle (2pt);
			\filldraw[color=black, fill=skyblue] (1pt, 0pt) circle (2pt);
			\filldraw[color=black, fill=skyblue] (1pt, 1pt) circle (2pt);
			\filldraw[color=black, fill=skyblue] (3pt, 3pt) circle (2pt);
			\filldraw[color=black, fill=skyblue] (6pt, 5pt) circle (2pt);
			\filldraw[color=black, fill=skyblue] (14pt, 12pt) circle (2pt);
			\filldraw[color=black, fill=skyblue] (28pt, 24pt) circle (2pt);
			\filldraw[color=black, fill=skyblue] (56pt, 63pt) circle (2pt);
			\filldraw[color=black, fill=skyblue] (140pt, 129pt) circle (2pt);
			\filldraw[color=black, fill=skyblue] (280pt, 280pt) circle (2pt);
			
			\matrix [below left] at (current bounding box.north east) {
				\node [bigskybluenode, label=right:CoSky «~SQL~» avec 3 attributs] {}; \\
			};
		\end{tikzpicture}
	}
	\caption{Temps de réponse de CoSky en SQL avec 3 attributs}\label{fig:temps_de_reponse_de_CoSky_en_sql_avec_3_dimensions}
\end{figure}
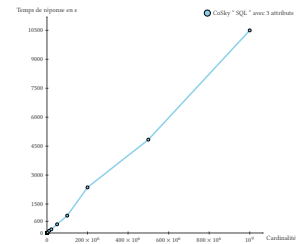

Pour l'évaluation de l'implémentation SQL de CoSky avec 3 dimensions, nous considérons la figure~\ref{fig:temps_de_reponse_de_CoSky_en_sql_avec_3_dimensions}. Les évaluations sont effectuées pour des cardinalités allant jusqu'à 1 milliard de tuples avec 3 dimensions. Dans le pire cas, le temps de réponse est de moins de $3$ heures. De plus, la lecture de la figure~\ref{fig:temps_de_reponse_de_CoSky_en_sql_avec_3_dimensions} semble nous indiquer qu'à nombre de dimension constant, l'évolution du temps de réponse est linéaire par rapport à l'augmentation de la cardinalité. Cela constitue une propriété très avantageuse de l'implémentation de CoSky en SQL, et c'est sans aucun doute la seule solution évaluée lors des expérimentations dans ce cas. 

\section{Conclusion}

Dans cet article, nous avons présenté des nouvelles méthodes efficaces de classement de Skyline.

La première proposée consiste en l'amélioration de la méthode dp-idp par l'utilisation d'une hiérarchie de dominance, offrant un classement plus rapide que la méthode initiale. Un exemple d'implémentation algorithmique de la méthode a été proposé.

La deuxième est la méthode CoSky, basée à la fois sur l'approche TOPSIS issue de l'aide à la décision multicritère et la mesure de similarité cosinus de Salton issue de la recherche d'information. Un exemple d'implémentation SQL ainsi qu'un exemple d'implémentation algorithmique de la méthode ont également été proposés.

L'algorithme DeepSky a été introduit afin de trouver les $k$ points de Skyline les mieux classés, c'est-à-dire ayant les scores les plus élevés, en utilisant le principe de Skyline multiniveaux couplé à la méthode CoSky.

Au vu de leur pertinence et de leur performance mises en évidence lors de l'évaluation expérimentale, nous pensons que les solutions exposées pourraient faire l'objet de futures publications, et d'un développement prochain d'une plateforme algorithmique de recherche \emph{open source}.

\section{Annexe}

\subsection{Préparation et remarques}\label{ssec:etape_0_preparation_et_remarques_prealables}

\subsubsection{Conversion d'une préférence Skyline}\label{sssec:conversion_d_une_preference_skyline}

Soit $r$ une relation composée des attributs $A_1, \dotsc,  A_m$. La conversion de $Pref(A_i) = MIN$ à $Pref(A_i) = MAX$, ou sa réciproque, est délicate dans les schémas d'ordonnacement basés sur un modèle vectoriel comme c'est le cas pour CoSky avec le cosinus de Salton. Il vaut mieux pour cela privilégier l'inversion mathématique, et non le complémentaire (pas plus qu'une conversion de minimum vers maximum ou de maximum vers minimum), des valeurs de l'attribut du Skyline.

En effet, le plus classiquement, la formule de calcul de valeurs complémentaires employée est $\forall t \in r, t[A_i'] = \bigvee A_i - t[A_i]$ où $\bigvee A_i$ est le \emph{supremum} de $A_i$, une valeur théorique suffisamment grande pour garantir que toutes les valeurs transformées restent positives et interprétables. $\bigvee A_i$ peut être la valeur maximale actuelle (\emph{optimum}) ou possible (maximum) de l'attribut $A_i$. Une approche alternative est d'utiliser la formule $\forall t \in r, t[A_i'] = \bigwedge A_i + \bigvee A_i - t[A_i]$ où $\bigwedge A_i$ est l'\emph{infimum} de $A_i$. Avec cette approche, \emph{infimum} et \emph{supremum} sont le plus souvent respectivement le minimum et les maximum.

Dans tous les cas, $\bigvee A_i$ (respectivement $\bigwedge A_i$) peut être inconnu ou varier.

L'inversion de valeur pose moins de problème, en particulier pour des valeurs strictement positives. Cette solution a plusieurs avantages et convient souvent mieux que l'utilisation des valeurs complémentaires. Cette méthode est indépendante de valeurs de bornes, il n'est ainsi pas nécessaire de connaître ou de calculer de \emph{supremum}, et maintient les proportions correctes. 

Cependant, même ainsi nous n'obtenons pas une correspondance parfaite à cause de la dispersion (l'étendue, la variance, la déviation absolue moyenne, la somme, etc.) des données.

\begin{example}
	Avec la relation \texttt{Pokémon} (cf. tableau~\ref{tab:relation_exemple}), si nous cherchons à convertir la condition de taux de \texttt{Victoire} le plus fort possible par celle du taux d'\texttt{Échec} (ou \texttt{Victoire}$^{-1}$) le plus bas possible, nous pouvons inverser les valeurs correspondantes. De la sorte, nous obtenons le tableau~\ref{tab:relation_exemple_3}, qui, pour des raisons de commodité, ne conserve aucun attribut ou commentaire, seulement le \texttt{RowId} des tuples. Notons cependant, bien que les relations de dominations entre les tuples restent correctes après transformation, les sommes des valeurs des attributs des points du Skyline (de \texttt{RowId} $1$, $2$ et $4$), respectivement $\Sigma = 160$ et $\Sigma = 9 / 140 \approx 0,0643$ sont différentes. Dans cette situation, la plupart des normalisations, comme celle par la somme employée par CoSky, donnent des résultats différents suivant que $Pref(\texttt{Victoire}) = MAX$ ou $Pref(\texttt{Échec}) = MIN$.
\end{example}

\subsubsection{Unification des préférences Skyline}\label{sssec:unification_des_preferences_skyline}

Nous considérons que, pour les schémas d'ordonnacement basés sur un modèle vectoriel comme c'est le cas pour CoSky avec le cosinus de Salton, il y a nécessité d'avoir des valeurs comparables entre elles, et donc d'avoir des préférences Skyline identiques. En effet, par essence, pour que le cosinus de Salton donne une mesure significative de similarité, il est crucial que les composantes des vecteurs représentent des entités comparables ou au moins normalisées de manière cohérente. Cela signifie que les valeurs des vecteurs doivent être sur des échelles comparables et avoir des unités similaires.

De la sorte, nous pouvons calculer un score par «~l'idéal mimimum~» ou alors un score par «~l'idéal maximum~», au choix, bien que nous préconisons, pour une meilleure précision et une économie de calculs, d'unifier les préférences en fonction de la préférence initialement majoritaire.

Le processus d'unification des préférences Skyline doit naturellement être effectué préalablement à toute étape de calcul.

\begin{figure}[htbp]
	\begin{minipage}{\linewidth}
		\centering
		\resizebox{.45\textwidth}{!}{
			\begin{tikzpicture}[
				line join=bevel
				]
				
				\draw [-stealth] (0pt, 0pt) -- (150pt, 0pt) node[anchor=north west] {$x$};
				\draw [-stealth] (0pt, 0pt) -- (0pt, 150pt) node[anchor=south east] {$y$};
				
				\draw [-stealth] (0pt, 0pt) -- (30pt, 30pt);
				\draw [-stealth] (0pt, 0pt) -- (120pt, 30pt);
				
				\draw[skyblue, line width=2pt] (30pt, 120pt) -- (60pt, 100pt) -- (90pt, 80pt) -- (120pt, 30pt);
				
				\filldraw[purple] (30pt, 30pt) circle (2pt) node[anchor=west]{$I+ (30, 30)$};
				\filldraw[black] (0pt, 0pt) circle (2pt) node[anchor=north]{$O$};
				
				\filldraw[black] (30pt, 120pt) circle (2pt) node[anchor=south]{$A (30, 120)$};
				\filldraw[black] (60pt, 100pt) circle (2pt) node[anchor=west]{$B$};
				\filldraw[black] (90pt, 80pt) circle (2pt) node[anchor=west]{$C$};
				\filldraw[cyan] (120pt, 30pt) circle (2pt) node[anchor=west]{$D (120, 30)$};
				
				\filldraw[black] (80pt, 110pt) circle (2pt) node[anchor=west]{$E$};
				\filldraw[black] (110pt, 70pt) circle (2pt) node[anchor=west]{$F$};
				\filldraw[black] (110pt, 100pt) circle (2pt) node[anchor=west]{$G$};
				
				\coordinate (D) at (120pt, 30pt);
				\coordinate (O) at (0pt, 0pt);
				\coordinate (I) at (30pt, 30pt);
				\pic [draw, ->, "$\theta$", angle eccentricity=1.5] {angle = D--O--I};
				
				\matrix [below left] at (current bounding box.north east) {
					\node [label=right:$\theta \approx 0.534~rad$] {}; \\
				};
			\end{tikzpicture}
		}
		\resizebox{.45\textwidth}{!}{
			\begin{tikzpicture}[
				line join=bevel
				]
				
				\draw [-stealth] (0pt, 0pt) -- (150pt, 0pt) node[anchor=north west] {$x$};
				\draw [-stealth] (0pt, 0pt) -- (0pt, 150pt) node[anchor=south east] {$y$};
				
				\draw [-stealth] (0pt, 0pt) -- (30pt, 80pt);
				\draw [-stealth] (0pt, 0pt) -- (120pt, 80pt);
				
				\draw[skyblue, line width=2pt] (30pt, 0pt) -- (60pt, 20pt) -- (90pt, 40pt) -- (120pt, 80pt);
				
				\filldraw[purple] (30pt, 80pt) circle (2pt) node[anchor=west]{$I+ (30, 80)$};
				\filldraw[black] (0pt, 0pt) circle (2pt) node[anchor=north]{$O$};
				
				\filldraw[black] (30pt, 0pt) circle (2pt) node[anchor=north]{$A (30, 0)$};
				\filldraw[black] (60pt, 20pt) circle (2pt) node[anchor=west]{$B$};
				\filldraw[black] (90pt, 40pt) circle (2pt) node[anchor=west]{$C$};
				\filldraw[cyan] (120pt, 80pt) circle (2pt) node[anchor=west]{$D (120, 80)$};
				
				\filldraw[black] (80pt, 10pt) circle (2pt) node[anchor=west]{$E$};
				\filldraw[black] (110pt, 50pt) circle (2pt) node[anchor=west]{$F$};
				\filldraw[black] (110pt, 20pt) circle (2pt) node[anchor=west]{$G$};
				
				\coordinate (D) at (120pt, 80pt);
				\coordinate (O) at (0pt, 0pt);
				\coordinate (I) at (30pt, 80pt);
				\pic [draw, ->, "$\theta'$", angle eccentricity=1.5] {angle = D--O--I};
				
				\matrix [below left] at (current bounding box.north east) {
					\node [label=right:$\theta' \approx 0.611~rad$] {}; \\
				};
			\end{tikzpicture}
		}
		\caption{Préf. $(\texttt{MIN}, \texttt{MIN})$ (gauche) / $(\texttt{MIN}, \texttt{MAX})$ (droite)}\label{fig:cosinus_de_salton_avec_preferences_skyline_unifiees_min_min}
	\end{minipage}
\end{figure}
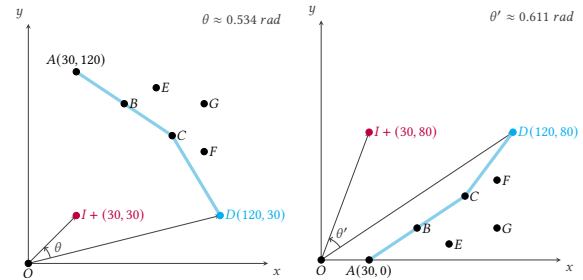

\begin{example}
	Dans la figure~\ref{fig:cosinus_de_salton_avec_preferences_skyline_unifiees_min_min}, nous considérons, par commodité de représentation, deux critères d'évaluation ayant, dans le premier cas, des préférences Skyline unifiées $(\texttt{MIN}, \texttt{MIN})$, et dans le second cas, des préférences Skyline mixtes $(\texttt{MIN}, \texttt{MAX})$~: le critère de préférence $\texttt{MIN}$ étant en abscisse, et celui de préférence $\texttt{MAX}$ en ordonnée.
	
	Les points $A$, $B$, $C$ et $D$ ne sont dominés par aucun autre point. Alors que les points $E$, $F$ es $G$ ne sont pas sur la frontière, ou front, (d'efficacité) de Pareto (l'ensemble des segments en couleur) parce qu'ils sont dominés par les autres points. On qualifie $A$, $B$, $C$ et $D$ d'efficaces, et de Pareto-optimaux. Le point idéal théorique, ou abstrait, noté $I+$, de coordonnées $(30, 30)$ à gauche de la figure~\ref{fig:cosinus_de_salton_avec_preferences_skyline_unifiees_min_min}, et de coordonnées $(30, 80)$ 	
	à droite, domine tous les points du Skyline.
	
	Le cosinus de Salton respectivement de l'angle $\theta$ et $\theta'$, tout deux formés par le vecteur allant de l'origine à $I+$ et le vecteur allant de l'orgine vers $D$, a pour valeur environ $0.534~rad$ à gauche de la figure~\ref{fig:cosinus_de_salton_avec_preferences_skyline_unifiees_min_min}, et une valeur différente, d'environ $0.611~rad$ à droite.
	
	Pour passer d'un cas à l'autre, nous avons employé le calcul de valeurs complémentaires classique (avec l'utilisation du \emph{supremum}) par raison de commodité. En effet, notamment, l'inversion d'uniquement certaines valeurs rend la représentation peu claire (certaines étant plus petites que $1$ là où d'autres sont bien plus grandes), mais quelle que soit la méthode de conversion employée, la problématique reste la même.
\end{example}

\begin{example}
	Le plus judicieux est donc d'unifier les préférences Skyline de la relation \texttt{Pokémon} (cf. tableau~\ref{tab:relation_exemple}), et de le faire avec $(\texttt{MIN}, \texttt{MIN}, \texttt{MIN})$ en inversant les valeurs de l'attribut \texttt{Victoire} (devenant ainsi \texttt{Victoire$^{-1}$} ou \texttt{Échec}), comme avec le tableau~\ref{tab:relation_exemple_3}.
\end{example}

\subsubsection{Normalisation et standardisation}\label{sssec:normalisation_et_standardisation}

La normalisation et la standardisation sont des techniques utilisées pour transformer les données afin qu'elles soient sur une échelle comparable, généralement entre $0$ et $1$. Nous présentons, dans ce paragraphe, quelques méthodes courantes de normalisation, avec leurs formules de calcul permettant de transformer une valeur originale $x$ en sa valeur normalisée ou standardisée $x'$.

\paragraph{Normalisation statistique}\label{para:normalisation_statistique}

La normalisation statistique, ou normalisation par plage, est adaptée pour des données sans distribution normale. C'est une méthode sensible aux valeurs aberrantes, idéale pour des données avec des minimums et des maximums bien définis. Sa formule de calcul est~:
\[x' = \frac{x - x_{min}}{x_{max} - x_{min}}\]

\paragraph{Standardisation}\label{para:Standardisation}

La standardisation centre les données autour de la moyenne et les met à l'échelle en termes d'écart-type. Elle est adaptée lorsque la distribution des données est approximativement normale. Cette méthode est moins sensible aux valeurs aberrantes que ne l'est la normalisation statistique. Avec $\bar{x}$ la moyennes des valeurs et $\sigma$ leur écart-type, sa formule de calcul est~:
\[x' = \frac{x - \bar{x}}{\sigma}\]

\paragraph{Normalisation par la mise à l'échelle décimale}\label{para:normalisation_par_la_mise_a_l_echelle_decimale}

La normalisation par la mise à l'échelle décimale transforme les données en déplaçant la virgule décimale des valeurs originales. Bien que simple à comprendre et à appliquer, cette méthode est moins couramment utilisée que les autres. Avec $k$ le plus grand nombre de chiffres à gauche de la virgule de $x$, sa formule de calcul est~:
\[x' = \frac{x}{10^k}\]

\paragraph{Normalisation par la somme}\label{para:normalisation_par_la_somme}

La normalisation par la somme transforme les données en divisant chaque valeur par la somme totale de toutes les valeurs. Cette méthode est particulièrement adaptée lorsque l'échelle relative des données est plus importante que leurs valeurs absolues, et elle convient bien pour les distributions proportionnelles. C'est cette solution que nous avons choisi pour la méthode CoSky. Avec $\Sigma x$ la somme de toutes les valeurs, sa formule de calcul est~:
\[x' = \frac{x}{\Sigma x}\]

\paragraph{Comparaison des méthodes}\label{para:comparaison_des_methodes}

Chacune des méthodes de normalisation ou de standardisation a ses applications privilégiées et ses avantages, et la méthode la plus appropriée et efficace dépend souvent de la nature spécifique des données et des objectifs de l'analyse.
	
	
	\bibliographystyle{ACM-Reference-Format}
	\bibliography{biblio}
	
	\InputIfFileExists{tex_append/annexes-fr}{}{}
\end{document}